\documentclass[11pt,a4paper]{article}
\usepackage{amsmath,geometry}
\usepackage{amssymb}
\usepackage{color}
\usepackage{graphicx}
\usepackage[sc,small,margin=1cm]{caption}

\newcommand{\be}{\begin{equation}}
\newcommand{\ee}{\end{equation}}
\newcommand{\bea}{\begin{eqnarray}}
\newcommand{\eea}{\end{eqnarray}}

\usepackage{hyperref}
\hypersetup{
    colorlinks=true,
    linkcolor=black,
    citecolor=black,
    filecolor=black,
    urlcolor=blue,
}
\usepackage{cite}

\begin{document}
\title{\vspace{1cm}\textbf{Manifest Modular Invariance\\ in the Near-Critical Ising Model} \\ \ \\} 
\author{\hspace{-0cm}Marcus Berg$^{1,2}$\\[5mm]
\hspace{-0cm} $^{1}$
{\it Department of Physics, Karlstad University,
 SE-651 88 Karlstad, Sweden} \ \\[5mm]
\hspace{-0cm} $^{2}$
{\it Nordita}\\
{\it Stockholm University and KTH Royal Institute of Technology}\\{\it Hannes Alfv\'ens v\"ag 12, SE-106 91 Stockholm, Sweden}\date{} }

\maketitle{}
\vspace{-10cm}
\hfill {\tt NORDITA 2022-174}

\vspace{10cm}
\begin{abstract}
Using recent results in mathematics, I point out 
that free energies and scale-dependent central charges away from criticality
can be represented in compact form where modular invariance is manifest. 
The main  example is the near-critical Ising model on a thermal torus, but the methods
are not restricted to modular symmetry, and apply to automorphic
symmetries more generally. One application is finite-size effects.
\end{abstract}

\newpage

\tableofcontents

\section{Introduction}

Modular invariance is often studied together with conformal invariance, but modular invariance is in itself a more basic property of any theory in two dimensions defined with periodic boundary conditions. In the standard representation of the torus as a lattice in the complex plane, physical quantities should not depend on the arbitrary choice of lattice basis. The group of modular transformations is generated by the S and T transformations, that correspond to switching lattice basis vectors (S transformation) and shifting one of the basis vectors by the other (T transformation). 

Physical quantities {\it should} not depend on arbitrary choices, like picking a gauge, or a frame of reference. However, sometimes for practical reasons one makes a choice, and then attempts to provide evidence that this choice did not matter. This is how the  Ising model on the torus away from its critical temperature has been treated in the literature. 
 
Alternatively, invariance can be built into the calculation from the beginning, here called ``mani\-fest'' invariance.  In gauge theory, working exclusively with gauge invariant quantities makes gauge invariance manifest. This effectively means that it has been proven prior to the calculation of interest that the method guarantees invariance, so the proof does not need to be revisited, as long as the assumptions are adhered to (i.e.\ no quantities have hidden gauge dependence). 

The same applies to modular invariance:  it can be shown using 19th century mathematics that the partition function $Z(\tau)$ of the critical Ising model is invariant under the modular group generated by S and T, so any quantity expressed in terms of $Z(\tau)$ is also invariant. The relevant mathematics is the theory of Jacobi theta functions, subsumed into one of its 20th century successors, such as the theory of Jacobi forms \cite{Eichler:1985}. 

If we introduce a mass scale by deviating from the critical point, the dispersion relation becomes that of a massive field, with the square root familiar from the Dirac or Klein-Gordon equations. The original proof no longer applies; ordinary Jacobi forms contain no square roots. The naive interpretation of this situation is that invariance is completely broken. But by the first paragraph of this introduction, that cannot be: physical quantities in the massive theory still cannot depend on the arbitrary choice of lattice basis. In the literature it is shown by perturbative expansion in the mass to lowest nontrivial order that the series coefficients can be expressed in the partition functions of various known massless theories. This is good evidence for modular invariance also in the massive case, but a low-order perturbative statement is not a general proof. More importantly, with this strategy invariance must be checked on a case by case basis and in each case seems to occur ``by accident''. 

The purpose of this paper is  to point out that more recent results in mathematics such as \cite{Berg:2019jhh} allow straightforward generalization of classical Jacobi forms (that can be called ``massless'', in this context) to ``massive Jacobi forms'', with the square roots built in from the beginning. By using those objects instead of the classical objects, which to a working physicist amounts to replacing some powers by some Bessel functions, we can now rely on the mathematics literature for the proofs, and claim ``manifest'' invariance in the near-critical Ising model, in the same sense as in the critical Ising model. 

The methods also apply more broadly, not just to other models defined on a torus, but to symmetries also in other dimensions, or without periodic boundary conditions. This parallels the generalization of modular forms, based on the the Lie group SL(2), to automorphic forms, that exist for any Lie group, or more general symmetries like
Kac-Moody algebras.

A short plan of this work is as follows. In section 2, the Ising model  away from criticality in general is reviewed,
including in section 3 how to check the deformed partition function using conformal perturbation theory. In section 4, the mathematical background is reformulated from a physics point of view, with details and proofs in the appendices.
With this as a new starting point, section 5 and 6 recalculates known quantities in a way that naturally leads to section 7, which contains the main new result of this paper: a mass expansion that is manifestly modular invariant at any order, in eq.\  \eqref{muexp0} together with eq.\ \eqref{derivatives}. However, as explained above, 
framing the mathematical results in the context of statistical physics, as done in sections 4,5 and 6, is as central an objective of this paper as the specific result in section 7. Section 8 outlines how this is relevant to finite-size effects.

Here are a few comments on related work. In \cite{DiFrancesco:1987ez} there is a computation of 
correlation function on the torus, extended to the Ashkin-Teller model in \cite{Saleur:1987tn}. In \cite{Duplantier:1987sd}, a comparison is made with lattice regularization, and the large-mass limit is considered. Perhaps most relevant for this paper, a few papers on finite-size effects are reviewed in section \ref{finite}.
A recent paper that relates high- and low-temperature expansions is \cite{Kostov:2022pvy}.
The paper \cite{Downing:2023uuc}, that constructs  tensor currents  in the two-dimensional theory of free massive fermions, was communicated to me by the authors.
Since the present paper uses the formulation of the Ising model as free massive fermions as the main example, it will be interesting to see if the results of \cite{Downing:2023uuc}
can be expressed usefully in  language more closely aligned with that used here.

\subsection{Physical mass scales and the Jacobi group}
\label{physicalmass}

Before introducing the example of the Ising model, here are a few general remarks to make the statements in the introduction more precise.
One way to state the issue is that  {\it the parameter $m$ in the differential operator $\partial_z\bar{\partial}_z + m^2 $ is not modular invariant}, whereas the combination $\mu=m^2\tau_2$ is,
where $\tau_2={\rm Im}\;  \tau$.

\begin{figure}[h]
\begin{center}
\includegraphics[width=0.5\textwidth]{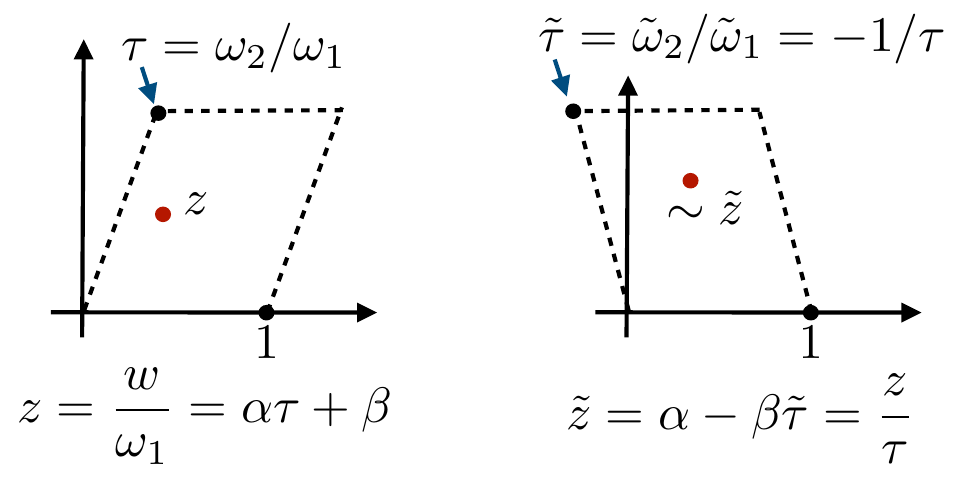}
\caption{A point $w=\alpha \omega_2+\beta\omega_1$ in terms of $\tau=\omega_2/\omega_1$ is $z=\alpha\tau+\beta$. Switching the lattice basis vectors as $\tilde{\omega}_1=\omega_2$ and  $\tilde{\omega}_2=-\omega_1$ (S modular transformation) makes $\tilde{\tau}=-1/\tau$ and $\tilde{z}=z/\tau$.
Here $\sim \tilde{z}$ means shifted by a lattice vector from $\tilde{z}$.}
\label{fig:ztr}
\end{center}
\end{figure}

To see this, first a quick review (for a pedagogical introduction, see Ch.1\ of \cite{Apostol}). Consider a torus lattice generated by complex numbers $\omega_1$ and $\omega_2$. Modular transformations $\Gamma$ act on this lattice as 
\be
\begin{pmatrix} \tilde{\omega}_1 \\ \tilde{\omega}_2 \end{pmatrix}
=  \begin{pmatrix}d & c  \\ b& a \end{pmatrix}
\begin{pmatrix} \omega_1 \\ \omega_2 \end{pmatrix} \label{Gammadef}
\ee
for integers $a,b,c,d$, where $ad-bc=1$.
It is convenient to rotate and rescale the lattice so the new first basis element is 1,
then the lattice is parametrized by the torus parameter $\tau = \omega_2/\omega_1$.
The modular transformation $\Gamma$ above acts on  $\tau$ as $\Gamma: \tau\rightarrow (a\tau+b)/(c\tau+d)$. 
A marked point $z$ on the torus transforms under $\Gamma$
as $z\rightarrow z/(c\tau+d)$. This amounts to\footnote{The overall sign
is fixed as follows.  Without the marked point $z$, we identify under the overall sign flip
$(a, b, c, d) \rightarrow -(a, b, c, d)$, since  it has no effect on $(a\tau+b)/(c\tau+d)$. In the standard theory of Jacobi forms  \cite{Eichler:1985}, this freedom is fixed as $z \rightarrow z/(c\tau+d)$. This action {\it is} sensitive to an overall sign flip: there is no $a\tau+b$ in the numerator to compensate.  } the simple statement that the {\it coordinates} of a marked point $z$ must change under $\Gamma$
as in fig.\ \ref{fig:ztr}.
The joint action on $\tau$ and $z$ is the action of the {\it Jacobi group} \cite{Eichler:1985}.

For example, the modular S transformation is represented on the lattice basis as $\tilde{\omega}_1=\omega_2$ and  $\tilde{\omega}_2=-\omega_1$, as in figure  \ref{fig:ztr}. To get from the left to the right panel in the figure, one can multiply by $\omega_1$ to go back to the $(\omega_1,\omega_2)$ basis, 
then transform to $\tilde{\omega}_1=\omega_2$ and  $\tilde{\omega}_2=-\omega_1$, then scale $\tilde{\omega}_1$ to 1  by dividing the ``tilde coordinates'' by $\tilde{\omega}_1$.
Note that although the lattices generated by $(\omega_1,\omega_2)$ 
and $(\tilde{\omega}_1,\tilde{\omega}_2)$ are the same, by definition of modular transformations,
  the area in the figure is changed
from Im $\tau$ in the left panel to Im $\tilde{\tau}$ in the right panel. 
As illustrated in figure  \ref{fig:ztr}, S  also transforms a
marked point $z$ on the torus to $\tilde{z} =  z/\tau$.

Here is a consequence of these simple observations. For a massive scalar field theory, consider the differential operator
\be  \label{noninv}
\partial_z\bar{\partial}_z + m^2 \; .
\ee
Now consider a general modular transformation, not just the S transformation.
As stated above,  $z\rightarrow z/(c\tau+d)$, so $\partial_z\bar{\partial}_z \rightarrow \partial_z\bar{\partial}_z |c\tau+d|^2$.
To be able to add $m^2$ and  $\partial_z\bar{\partial}_z$, they must transform the same way. This means $m^2$ transforms 
as $m^2\rightarrow m^2 |c\tau+d|^2$. 

The good news is that the imaginary part $\tau_2$ transforms under the Jacobi group as $\tau_2 \rightarrow \tau_2/|c\tau+d|^2$,
which is the opposite of how eq.\  \eqref{noninv}  transforms.
So we can easily make eq.\ \eqref{noninv} invariant under the Jacobi group by simply multiplying by $\tau_2$. Indeed, 
the differential operator
\be \label{jacobigroup}
\tau_2\partial_z\bar{\partial}_z + m^2\tau_2
\ee
is invariant under the Jacobi group. 

From a physics point of view, using \eqref{jacobigroup} instead of \eqref{noninv} is a minor modification.
An invariant quantity like the partition function can be series expanded in the combination $\mu:=m^2\tau_2$ in \eqref{jacobigroup}, and each term will be separately invariant. Physically,
this can be understood as saying that if the mass measured by an observer
on the torus is expressed in units of inverse area of the torus, that number is invariant
if we relabel the torus lattice, as it should if that relabelling has no physical meaning.

To be clear, the statement that $m$ transforms if defined as in \eqref{noninv} does {\it not} mean that the physical Ising model away from criticality somehow fails to be modular invariant. 
As was argued in the previous section, that would make no sense: the  mass represents the correlation length. The statement here is that 
the parameter $m$ in eq.\ \eqref{noninv}, which is sometimes also called mass, is not an invariant label characterizing the amount of deviation from criticality, whereas $\mu$ is.

\section{Mass deformation of the critical Ising model}
\subsection{Ising model}
The Ising model  away from criticality was formulated in a way useful for this analysis
in a beautiful 1987 paper by Saleur and Itzykson 
 \cite{Saleur1987}, the main starting point for this work. The model 
has correlation length $\xi = 1/|M|$, where 
\be
M \propto {T-T_{\rm c} \over aT_{\rm c}}
\ee
with $a$ the lattice spacing. In the scaling regime the theory
 is described by a massive fermion field $\psi$ on a torus ${\mathcal T}$:
\be   \label{Zising0}
Z_{\rm Ising} =\sum_{\rm b.c.} \int {\mathcal D}\psi  {\mathcal D} \bar{\psi}\, 
\exp\left(2\int_{\mathcal T} d^2 z\left( \psi \bar{\partial} \psi - \bar{\psi}\partial \bar{\psi} + m\bar{\psi}\psi \right) \right)
=\sum_{\rm b.c.} ({\rm det}(-\nabla^2 +M^2))^{1/2}
\ee
where $\nabla^2 = \tau_2 \partial_z \bar{\partial}_z$ and  $M^2=\tau_2 m^2$. (As in the previous section, note that $d^2 z$ is not invariant, but $d^2 z/\tau_2$ is.) 
 The  sum over topologically distinct boundary conditions \!(b.c.)\! has 4 terms,
 denoted $D_{\alpha,\beta}$ for $\alpha,\beta=0$ (periodic) and $1/2$ (antiperiodic):
\be  \label{Ising1}
Z_{\rm Ising} = {1 \over 2}\left(D_{{1 \over 2},{1\over 2}}(M)+D_{0,{1\over 2}}(M)+D_{{1 \over 2},0}(M)+D_{0,0}(M)\right) \; .
\ee
Here the functional determinants $D_{\alpha,\beta}$ are computed\footnote{In \cite{Saleur1987}, there is a different ordering of indices, so $D^{\rm there}_{0,{1 \over 2}}=D^{\rm here}_{{1 \over 2},0}$. In general there is an absolute value, but there as here,
each factor is real. See also appendix  \ref{invmu} for a comment on $D_{0,0}$.}  by $\zeta$-function regularization  in \cite{Saleur1987}:
\be \label{Dab}
D_{\alpha,\beta}(M) = e^{-\pi \tau_2 \gamma_{\alpha}(t)} \prod_{n=-\infty}^{\infty}\!\!
\left(1-e^{-2\pi i \beta-2\pi i \tau_1(n+\alpha)-2\pi \tau_2\sqrt{(n+\alpha)^2+t^2}}\right)
\ee
where $\tau_1 = {\rm Re}\, \tau$, $\tau_2 = {\rm Im}\, \tau$,
and the dependence on  the mass parameter $M$ is through
\be \label{tparam}
t ={ M|\omega_1| \over 2\pi}  \; ,
\ee
where as reviewed in the previous section, the torus lattice in the complex plane is specified by two complex numbers $\omega_1$ and $\omega_2$. From eq.\ \eqref{Gammadef},
we see $|\omega_1|\rightarrow |c\tau + d||\omega_1|$, so $t$ in \eqref{tparam} is not invariant. In eq.\ \eqref{Dab},  the expressions  $\gamma_{\alpha}(t)$ for the values $\alpha=0,1/2$ considered here are
\bea
\gamma_0(t) &=& {1 \over 6}-t+t^2\ln(4\pi e^{-\gamma}\sqrt{\tau_2/A})+{t^4 \over 2} \int_0^1 d\lambda (1-\lambda)\sum_{n=1}^{\infty}{1 \over (n^2+\lambda t^2)^{3/2}}   \\
 \gamma_{1/2}(t) &=&  -{1 \over 12}+t^2\ln(\pi e^{-\gamma}\sqrt{\tau_2/A})+{t^4 \over 2} \int_0^1 d\lambda (1-\lambda)\sum_{n=0}^{\infty}{1 \over ((n-1/2)^2+\lambda t^2)^{3/2}}  \; . 
\eea
with the Euler-Mascheroni constant $\gamma=0.5772 \ldots$ (the $\gamma$ without subscript). 

These somewhat awkward representations for $\gamma_{\alpha}(t)$ have the advantage that the sums inside the integrals clearly converge, whereas convergence is less clear if we first perform the integrals. Alternative representations will be introduced below.

The sum of terms in \eqref{Ising1} gives
\bea  \label{Zising}
Z_{\rm Ising} &=& 
{1\over 2}e^{-2\pi\tau_2 E_1}\Big(\sum_{\pm}\prod_{n=-\infty}^{\infty}\Big(1\pm e^{2\pi i (n+1/2)\tau_1-2\pi \tau_2\sqrt{(n+1/2)^2+t^2}}\Big)\\
&&\quad+e^{2\pi\tau_2 (E_1-E_2)}\sum_{\pm}\prod_{n=-\infty}^{\infty}\Big(1\pm e^{2\pi i n \tau_1-2\pi \tau_2\sqrt{n^2+t^2}}\Big)\Big)
\nonumber
\eea
where the zero-point energies are given by $E_1=\gamma_{1/2}/2$, $E_2=\gamma_{0}/2$, and to facilitate comparison with \cite{Saleur1987}, the combinations
appearing in \eqref{Zising} are
\bea \label{Ezero}
2E_1 &=& -{1 \over 12}+t^2\ln(\pi e^{-\gamma}\sqrt{\tau_2/A})+{1 \over 2}t^4 \int_0^1 d\lambda (1-\lambda)
\sum_{n=1}^{\infty}{1 \over ((n-1/2)^2+\lambda t^2)^{3/2}}  \label{Ediff}  \\
2(E_1-E_2) &=& -{1 \over 4}+t-t^2\ln 4+{1 \over 2}t^4 \int_0^1 d\lambda (1-\lambda)
\sum_{n=1}^{\infty}\left({1 \over ((n-1/2)^2+\lambda t^2)^{3/2}} -{1 \over (n^2+\lambda t^2)^{3/2}}\right)  \; .  \nonumber
\eea
As a check, at criticality $t=0$, eq.\ \eqref{Ezero}
gives $E_1=-1/24$ (antiperiodic), $E_2=1/12$ (periodic), as expected for fermions. 

The partition function $Z_{\rm Ising}$  in \eqref{Zising} was already found by Ferdinand and Fisher in 1969 \cite{Ferdinand:1969zz}.
In \cite{Saleur1987} a problem is noted: although $Z_{\rm Ising}$ in \eqref{Zising} should be modular invariant:
\be 
Z_{\rm Ising}(\omega_1,\omega_2,M) = Z_{\rm Ising}(\tilde{\omega}_1,\tilde{\omega}_2,M) \; ,
\ee
invariance is not manifest when $Z_{\rm Ising}$ is expressed in terms of $\tau$ and $t$.
One could argue that there is only scale invariance at the critical point $t=0$, so that working
with $\tau$ is not a good choice anyway.
More technically, one could be worried
that  square roots like $\sqrt{n^2+t^2}$ in \eqref{Zising} might invalidate the simple S transformation property of the partition function, compared to at criticality, where $t=0$.
In \cite{Saleur1987}, perturbative expansion in $M$ is used to argue for modular invariance
also in the sense of acting on $\tau$:
\be  \label{expansion}
\ln Z_{\rm Ising} = \ln Z_{1/2} + {M \sqrt{A} \over 2Z_1Z_{1/2}}+M^2 A \left({1 \over 4\pi}\ln\left(Z_1\sqrt{A}e^{\gamma} \over \pi\right)-{1 \over 4\pi} {\sum_{\alpha,\beta} D_{\alpha,\beta} \ln  D_{\alpha,\beta} \over Z_{1/2}} - {1 \over 8(Z_1Z_{1/2})^2}\right) + \ldots
\ee
where each term  only depends on $Z_{1/2}$, $Z_1$ (the fermion/boson partition functions at  $M=0$) and the parameters
$M$ and $A$. This gives evidence that $Z_{\rm Ising}$ is
modular invariant, by invariance of the $t=0$ ingredients. But this is no proof,
since  it only refers to the displayed terms in the expansion  in $M$.\footnote{This expression
differs between \cite{Saleur1987} and \cite{Itzykson1989} in the convention for terms involving $A$.}

Note that  \eqref{expansion} is (apart from the anomalous term with the logarithm)  an expansion in 
the dimensionless parameter $M^2 A$, with $A$ the area of the parallelogram spanned by $\omega_1$ and  $\omega_2$,
rather that  just an expansion in the mass parameter $M$ itself.
Expressed in $\tau=\omega_2/\omega_1$, 
the modular-invariant area is $A = |\omega_1|^2 \tau_2$, so $M^2  A =(2 \pi t)^2\tau_2$, where $t^2$ and $\tau_2$ each transform, but their product is invariant. We see that $t$ corresponds to the non-invariant $m$ parameter of the previous section, and the modular-invariant parameter $\mu=m^2 \tau_2$ used in this paper corresponds to the $M^2 A$ combination for the $(\omega_1,\omega_2)$ lattice, but here expressed intrinsically in the  $\tau$ frame.

The interested reader is referred to \cite{Saleur1987} for more details on eq.\ \eqref{expansion}, since it is not needed here: it is replaced by the simpler expansion eq.\ \eqref{muexp0} below.

Some readers may still wonder: is the $\tau$ frame really useful away from criticality, perhaps it would be better to stay in the $(\omega_1,\omega_2)$ lattice? For some questions that may indeed be the best choice. But ``near-critical'' in the title of this paper means that surely the powerful methods of Jacobi forms at criticality should  confer some advantage to  embedding them in a similar setting away from criticality. It could well be that far from criticality, this picture is not very useful.

\subsection{The running central charge}
\label{runningC}
In the $\tau_2\rightarrow \infty$ limit, we can neglect the exponentially suppressed terms in \eqref{Zising}:
\be  \label{ZisingLimit}
Z_{\rm Ising} \rightarrow
e^{-2\pi\tau_2 E_1}\Big((1+\ldots)+e^{2\pi\tau_2 (E_1-E_2)}(1+\ldots)\Big)
\rightarrow e^{-2\pi\tau_2 E_1}
\ee
as long as $E_1-E_2<0$, which by  the second equation in \eqref{Ediff} holds for $t\leq 1/4$, if 
terms of order $t^2$ and higher are neglected, i.e.\ the theory is not far from criticality. 
Close to $t=0$ we have $E_1<0$, so $Z_{\rm Ising}$ diverges as $\tau_2\rightarrow \infty$.
The zero-point energy determines the {\it running central charge} 
as used in \cite{Saleur1987,Itzykson1989}\footnote{The definition differs between \cite{Saleur1987} and \cite{Itzykson1989} by an overall factor
of 12. }
\be \label{Corig}
C(t) = \lim_{\tau_2\rightarrow \infty}{\ln Z(t) \over 2\pi \tau_2}  = -E_1(t) \; .
\ee
If $A$ is viewed as fixed, the logarithmic term $\ln\sqrt{\tau_2/A}$ in $E_1$ in eq.\ \eqref{Ezero} dominates as $\tau_2\rightarrow \infty$,
as it must for a divergent specific heat. 
%At a given renormalization scale, the logarithmic term vanishes. 
%The renormalization point used here is $|\omega_1|=2\pi$,
%then $A=(2\pi)^2\tau_2$ and the combination $\ln (2\pi \sqrt{\tau_2/A})=0$. 
The arbitrariness in the choice of renormalization point only affects terms of order $t^2$. 

In particular, the $t^4$ term in the expansion\footnote{In \cite{Saleur1987}, $C_0$ is related to the usual central charge
by  a factor of 12, due to the factor in the previous foonote. }
\be \label{Cexpansion}
C(t)=C_0 + C_2t^2 + C_4t^4+\ldots
\ee
 is finite and unambiguous:
\bea \label{C4}
C_4 =C(t)\Big|_{t^4}&=& 
%E_1(t)\Big|_{t^4} = -{1\over 2} \gamma_{1/2}(t)\Big|_{t^4} =
 -{t^4 \over 4}\int_0^1 d\lambda \, (1-\lambda)
\sum_{n=1}^{\infty}{1 \over ((n-1/2)^2+\lambda t^2)^{3/2}}\Big|_{t^4}  \nonumber\\
&=& \sum_{n=1}^{\infty}
\left({1 \over 2}\sqrt{(n-1/2)^2+t^2}-(n-1/2)-{t^2 \over 2n-1}\right)\Big|_{t^4} \\
& =& -\sum_{n=1}^{\infty}{t^4 \over
(2n-1)^3}  =  -{7 \over 8}\zeta(3)t^4 \; . \nonumber
\eea
A theme in this paper is to recalculate $C_4$ in several different ways.

The calculation \eqref{C4}  illustrates the comment above about convergence: after integration,
the sum of each of the three terms on the second line in \eqref{C4} individually diverge, only the total is finite, so convergence is not apparent. 
By contrast, the representation advocated in later sections of this paper is more explicit than an integral representation, but still convergent by inspection.

%\subsection{Running operator dimensions}
%Similarly to the running central charge, one can consider the running of operator dimensions  \cite{Saleur1987}. 
%In the massless conformal field theory, operator dimensions appear in the partition function as $q^{L_0}$, where $q=e^{2\pi i \tau}$.
%For states with $L_0$ eigenvalue $n$, the absolute value is $e^{2\pi \tau_2 E} e^{-2\pi \tau_2 (n+\alpha)}$. 
%In the massive theory with $t \neq 0$, this is deformed to $e^{2\pi \tau_2 E(t)} e^{-2\pi \tau_2 \sqrt{(n+\alpha)^2+t^2}}$.
%
%More precisely, the Hamiltonian for arbitrary scale $\omega_1$ can be written
%\be \label{H}
%H = {2 \pi \over \omega_1}\left(L_0 + \bar{L}_0-{C \over 12}\right)
%\ee
%with the basic scale set at $\omega_1=2\pi$ in the conventions of \cite{Saleur1987}. With mass, there is no clear distinction between left- and right-movers, and
%$C$ contains contributions from both.
%
%To first order in $t$ we have
%\be \label{D4}
%E_2(t) = -{1 \over 2}\gamma_0(t) = -{1\over 12}+{t \over 2}+\ldots
%\ee
%so if we call the coefficient of the  order $t^n$ term $D_n$, we see that 
%in particular the linear term has $D_1=1/2$. 
%Both this
%result and $C_4$ in \eqref{C4}
%from the previous subsection can be checked by a perturbative calculation, as in the next section.
%
%As a reminder,  there are only two different zero-point energies $E_1$ and $E_2$, but four different periodicities of fermions going around the cycles of the torus. This is because the energy only depends on the periodicity in one of the two directions. 

\section{Perturbative approach}
Previous sections concerned specifically the massive deformation of the critical Ising model. 
More generally, one can add to the Hamiltonian of any critical theory a perturbing operator $V$:
\be
V = {G \over 2\pi} \int_{\mathcal T} \! d^2z \; \varphi
\ee
where the field $\varphi$ has any weight $(h,h)$. It is useful to form the dimensionless
coupling constant
\be
g = G \left({|\omega_1|\over 2\pi}\right)^{2-2h} \; . 
\ee
We can view the mass in the previous section
as a perturbation of this kind, with $\varphi=\psi\bar{\psi}$, so $G=m$, $g=t$ and the weight $h=1/2$.
With this formalism we can also consider other perturbations, for example a magnetic field perturbation,
with $\varphi$ being the spin operator $\sigma$ with $h=1/16$, but this will not be explored in this work. 

The perturbed partition function (matrix element)\footnote{The  book \cite{Itzykson1989} has $x_1$ instead of $x_i$, this is a typo.} is
\be \label{Z0}
Z_0(g,\tau_2) = e^{\pi c \tau_2/6} \sum_{n=0}^{\infty}\left({g \over 2\pi} \right)^n \int_{\rho\leq |x_1|\leq \cdots \leq |x_n| \leq 1}
\left(\prod_{i=1}^{n} {d^2 x_i \over (x_i\bar{x}_i)^{1-h}}\right) \langle 0 | \varphi(x_1,\bar{x}_1)\cdots
\varphi(x_n,\bar{x}_n)|0\rangle \; . 
\ee
where the complex torus coordinates $z_i$ were transformed to complex sphere coordinates $x_i=e^{2\pi i z/\omega_1}$, and a cutoff $\rho$
around the origin in each $x_i$ plane was introduced.

As in \eqref{Corig}, a coupling-constant-dependent ``central charge'' $C(g)$ is defined from this as
\be  \label{defC}
C(g) = \lim_{\tau_2\rightarrow \infty}{\ln Z_0(g,\tau_2) \over 2\pi \tau_2} \; . 
\ee
The quantity $C(g)$ in \eqref{defC} can be computed in perturbation theory,
\be
C(g) = C_0 + g^2C_2 + g^4 C_4+\ldots
\ee
where $C_0=c/12$, the usual central charge.
For example, to compute the first correction $C_2$ we only need the 2-point function:
\be \label{corr}
  \langle 0 | \varphi(x_1,\bar{x}_1)\varphi(x_2,\bar{x}_2)|0\rangle = {1 \over |x_{12}|^{4h}}
\ee
where $x_{12}=x_1-x_2$. Inserting \eqref{corr} in \eqref{Z0}, we obtain the 2nd order term in  $Z_0(g,\tau_2)$ as
\be \label{Z02}
Z_0(g,\tau_2)\Big|_{g^2} =  {g^2 \over (2\pi)^2}\int_{\rho}^1 {d^2 x_1 \over |x_1|^{2-2h}} \int_{|x_1|}^1 {d^2 x_2 \over |x_2|^{2-2h} }{1 \over |x_1-x_2|^{4h}}
\ee
which gives (more details for interested readers in Appendix \ref{pert}):
\be \label{C2general}
C_2 =
 \sum_{n=0}^{\infty}{1 \over 2n+2h}\left({\Gamma(n+2h) \over n! \, \Gamma(2h)}\right)^2
=2^{-1-4h} {\Gamma(1/2-h)\Gamma(h) \over \Gamma(1-h)\Gamma(h+1/2)}
\ee
which converges for $h<1/2$, in particular for the spin operator $h=1/16$. For the mass operator
$h=1/2$, eq.\ \eqref{Z02} can be defined by a limiting procedure as $h\rightarrow 1/2$ from below.

In  \cite{Itzykson1989} it is stated that it would be preferable
to compute $C(g)$ without expanding in mass:\\[-5mm]

\hspace{2mm}
\parbox{15cm}{
\begin{center}
{\it Un progr\`es tr\`es int\'eressant serait de trouver des expressions compactes pour ces quantit\'es\\ au lieu d'un d\'eveloppement en puissances du couplage. \\Diverses indications sugg\`erent que cet espoir n'est pas exclu.}
\end{center}}

Such nonperturbative calculations
 could be feasible with the method presented below. The first step taken here is only to repackage existing results in a manifestly invariant way.

\subsection{Example: thermal perturbation and central charge}
We  come back to the mass operator as a perturbation away from $T=T_{\rm c}$.
The first coefficient in the running central charge $C_2$ is clearly divergent: \eqref{C2general} has a pole
as $h\rightarrow 1/2$. There is a finite term $C_{2,{\rm finite}}= \ln 2$, which we
can attempt to match to the finite piece at order $t^2$ in the previous section,
but  the value depends on the renormalization prescription. 

The 4th order coefficient in the $t$ expansion is finite and unambigious:
\be \label{C4val}
C_4  =
-2\int_{\rho}^1{dx_4 \over x_4^2}\int_{\rho}^{x_4}{dx_3 \over x_3^2}
\int_{\rho}^{x_3}dx_2 \int_{\rho}^{x_2}dx_1\sum_{n=0}^{\infty}\left({x_1 x_2 \over x_3 x_4}\right)^{2n} = 
 -{7 \over 8}\zeta(3) \; ,
\ee
which matches with the exact result \eqref{C4} expanded to this order. So this is a useful check of the perturbative prescription. 
To show \eqref{C4val} using
\eqref{Z0}, one needs
 the correlation function
\be \label{corr2}
 \langle 0 | \varphi(x_1,\bar{x}_1)\varphi(x_2,\bar{x}_2)\varphi(x_3,\bar{x}_3)\varphi(x_4,\bar{x}_4)|0\rangle_{\rm connected} = 
 -{1 \over x_{13}x_{24}\bar{x}_{14}\bar{x}_{23}} +\mbox{similar terms}
\ee
that follows from representing the correlation function 
as a Pfaffian of $1/x_{ij}$. Performing the integrals like in the 2-point case above gives for the $n$-point term:
\be \label{generalexp}
\sum_{n=2}^{\infty}
(-1)^{n+1}(2^{2n}-2) {\Gamma(n-{\scriptstyle {1 \over 2}}) \over 4\sqrt{\pi} n!}
\zeta(2n-1)t^{2n} = -{7 \over 8}\zeta(3)t^4 + {31 \over 16}\zeta(5)t^6
-{635 \over 128} \zeta(7) t^8 + {3577 \over 256 }\zeta(9)t^{10}+ \ldots
\ee
More details are given in \cite{Saleur1987}. The multiple integral in  \eqref{C4val}, leading 
to \eqref{generalexp}, will be reproduced by a single integral in  
\eqref{goodexp} below.

%
%\subsection{Example: thermal perturbation and operator dimensions}
%Similarly we can compute 
%\be
%D_1^{\rm mass} = 1/2
%\ee
%which matches with the corresponding calculation in \eqref{D4},
%from the deformed partition function. 

%\subsection{Example: spin operator}
%For completeness, we also briefly discuss perturbation by the spin operator $\sigma$. 
%For the ratio of expectation values of cumulants of the magnetization,  \cite{Saleur1987} finds
%\be
%R = {-C_4^{\rm spin} \over \pi (C_2^{\rm spin})^2} = -2.46048 \pm 0.00005 \ldots
%\ee
%Simulations give $R\approx 2.46044\pm 0.00002$. The first analytic approximation is $R\approx 31/(4\pi)$. 
%
%\subsection{Example: energy shifts}
%Another example is the ratio of the shift at order $g^2$ of the energies of the ground-state and first excited state:
%\be
%r = {C_2^{\rm spin} \over D_2^{\rm spin}} = -1.03926 \ldots
%\ee
%This concludes the review.

\section{Towards manifest modular invariance}

For the purposes here, the expression $Z_{\rm Ising}$ in \eqref{Zising} can
usefully be repackaged as follows. 
Consider each piece $D_{\alpha,\beta}$ from  \eqref{Dab} separately.
In \cite{Berg:2019jhh}, and references therein, we find the following definition of a deformed partition function:
\bea  \label{Zmuz}
Z_{\alpha,\beta,m}\left(\tau\right) &=& e^{-8 \pi  c_{\alpha,m}\tau_2}\prod_{n=-\infty}^{\infty}\prod_{\pm}\left(1- e^{\pm 2 \pi  i \beta-2\pi \tau_2 \sqrt{m^2 + \left( n \pm \alpha \right)^2}+ 2 \pi i \left(n \pm \alpha \right)\tau_1}\right), 
\eea
with
\begin{equation} \label{czero}
c_{\alpha,m} :=\frac1{(2\pi)^2} \sum_{\ell \geq 1} \cos( 2\pi \ell \alpha)\!\! \int_0^\infty e^{-\ell^2x-\frac{\pi^2m^2}{x}}dx=\frac{m}{2\pi} 
\sum_{\ell\ge 1}\cos(2\pi \ell \alpha) \frac{K_1\left(2\pi  \ell m\right)}{\ell} \; ,
\end{equation}
where $K_1$ is the standard K-Bessel function. Comparing with eq.\  \eqref{Dab}, the $D_{\alpha,\beta}$
seems almost like it could be expressed in  \eqref{Zmuz}, except there seem to be some additional signs $\pm$ in 
$e^{\pm 2\pi i \beta}$ and by $(n\pm \alpha)$  in \eqref{Zmuz}.
However, here we only consider zero or half-integer $\alpha$ and $\beta$. Then, $e^{\pm 2\pi i \beta}$ is independent of $\pm$. Also, the shift in 
$(n\pm 1/2)$ is inconsequential when we sum over all $n$. We then see
 that \eqref{Zmuz} consists of two copies of \eqref{Dab}, if the $c_{\alpha,m}$ and $\gamma_{\alpha}(t)$ agree between these two expressions for the partition function,
when relating $m$ and $t$. In section \ref{zerop},
 I review the textbook argument that $c_{\alpha,m}$ and $\gamma_{\alpha}(t)$ do match, and   in section \ref{modularcov}
this will allow repackaging of $D_{\alpha,\beta}$ from \eqref{Dab} in manifestly modular invariant form. 

\subsection{Zero-point energy}
\label{zerop}
The purpose in this subsection is to  compare the Bessel-sum representation
of $c_{\alpha,m}$ in  \eqref{czero} to the integral representations
of $\gamma_{\alpha}(t)$ in \eqref{Ezero}. Neither merits the precise name ``zero-point energy'', but both have
the same functional form as the zero-point energies $E_i$, and only differ by  an overall factor.  

It is explained for example in the textbook \cite{Mussardo:2020rxh}\footnote{there, $r=2\pi m$}, section 20.11, that the running central charge in a mass-deformed theory is
\be \label{cBessel}
c_{\pm}(m) = \mp {12 m\over \pi} \int_0^{\infty}
\!d\theta \,  \cosh \theta \log(1\mp e^{-2\pi m \cosh \theta}) =
{12  m \over \pi}\sum_{n=1}^{\infty} {(\pm1)^{n-1} \over n}K_1(2\pi n m) \; . 
\ee
(See also eq.\ (90), (91) and (99) of \cite{Klassen:1990dx}.)
This is 24 times $c_{\alpha,\mu}$ in \eqref{czero} and the two choices $\alpha = 0$ or 1/2  for the quasiperiodicities of the fermions correspond
to the $\pm 1$ in eq.\ \eqref{cBessel}.
In particular,  $c_+$ is related to $\gamma_0$, and $c_-$ is related to $\gamma_{1/2}$. 

\begin{figure}[h]
\begin{center}
\includegraphics[width=0.5\textwidth]{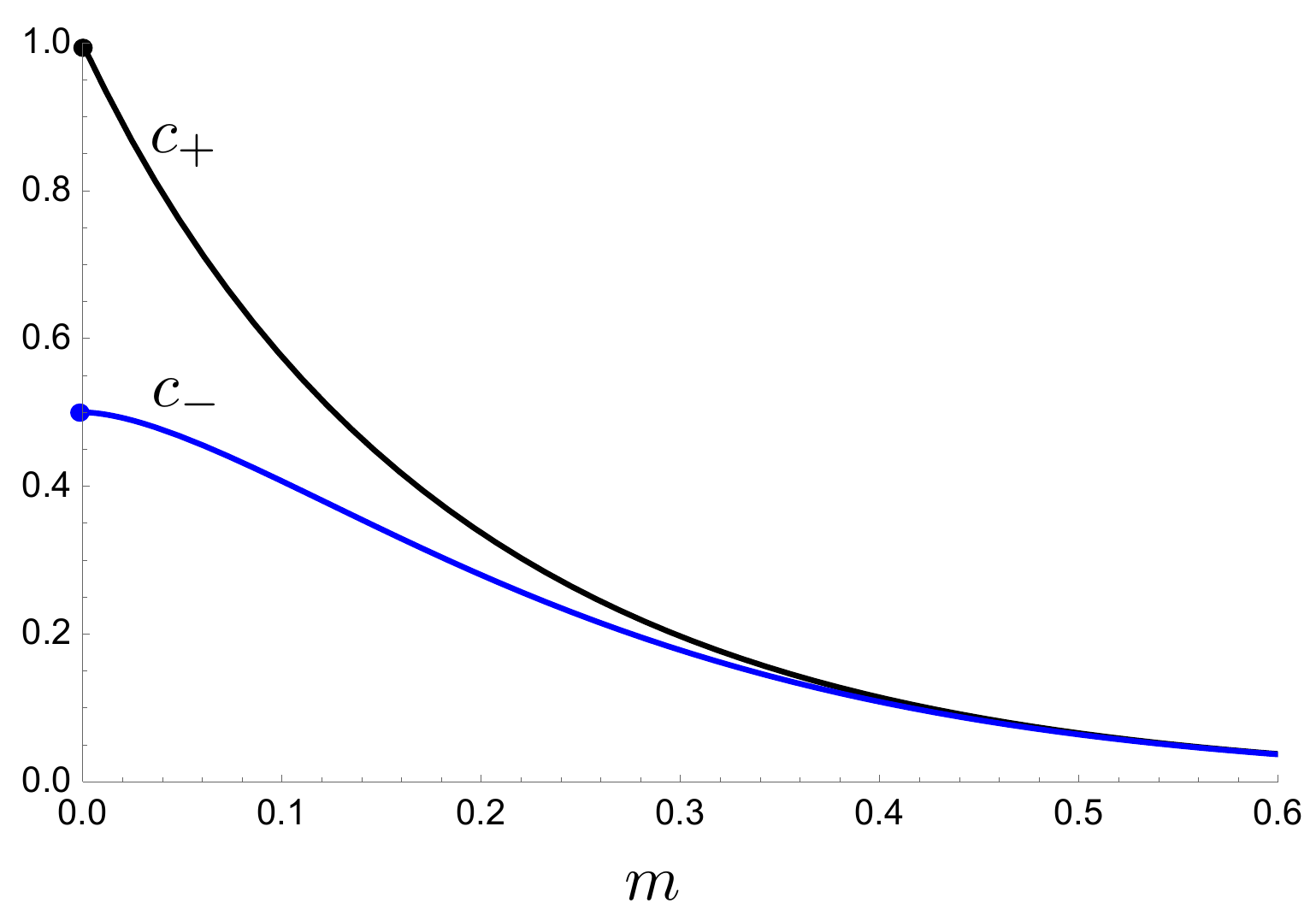}
\caption{Running central charges for two fermion quasiperiodicities as function of the mass parameter.}
\label{fig:c}
\end{center}
\end{figure}

The running central charges $c_+$ and $c_-$ are plotted in figure \ref{fig:c}.
It shows that $c_+ \rightarrow 1$ and $c_- \rightarrow 1/2$ as $m \rightarrow 0$. This normalization merits the use of  $c$ for running central charge. (Recall
from section \ref{runningC} that there is a factor of 12 between $C(t=0)$ and the usual central charge,
so there is an additional factor of 2 between  $c_{\alpha}$ and $|E_i|$.)

Now, it is shown in the textbook \cite{Mussardo:2020rxh} that $c_+$ can be rewritten as follows (with a similar expression for $c_-$:) 
\be  \label{cexpl}
c_+(m) = 1 -6m+ {6 m^2  }\left(-\log  m
+{1 \over 2}+\log (2) -\gamma\right) -{12 }\sum_{n=1}^{\infty}\left(\sqrt{n^2+m^2}
-n - {m^2 \over 2 n}\right) \; . 
\ee
%\be  \label{cexpl}
%c_+(r) = 1 -{3 r \over \pi} + {3 r^2 \over 2\pi^2}\left(\log {1 \over r}
%+{1 \over 2}+\log (4\pi) -\gamma\right) -{6 \over \pi}\sum_{n=1}^{\infty}\left(\sqrt{(2 n \pi)^2+r^2}
%-2n\pi - {r^2 \over 4 n \pi}\right)
%\ee

Individual terms in the sum over $n$ are divergent, 
but the sum as a whole can be truncated.
For the interested reader, a few details
how to show using elementary manipulations that $c_+$ in \eqref{cBessel} equals
\eqref{cexpl}, are given in appendix \ref{besselapp}. In particular, the sum over a square root can be viewed as arising from the fact that the Fourier transform
of a square root is a Bessel function.

A clear alternative exposition of this is given in \cite{Downing:2023uuc}, appendix D. 

The zero-point energies in eq.\ \eqref{Ezero} 
are  integral representations over a variable $\lambda$, where the sum over $n$ is manifestly convergent before integration.
Like in \eqref{C4}, the integral can be performed as\footnote{This is discussed in \cite{Duplantier:1987sd}. There is a typo on the left-hand side
of their equivalence, $t$ should be $t^2$.}
\be  \label{connect}
{t^4 \over 2}
\int_0^1 d\lambda(1-\lambda){1 \over (n^2+\lambda t^2)^{3/2}} =- \left(2\sqrt{n^2+t^2}-2n - {t^2 \over n} \right)
= {t^4 \over 4n^3} - {t^6 \over 8n^5} + {5 t^8 \over 64 n^7}+\ldots
\ee
where we recognize the sum in \eqref{cexpl}.
So together, eqs.\ \eqref{connect}, \eqref{cexpl} and \eqref{cBessel}  provide the connection between the Bessel sum in  \eqref{czero} and the
 zero-point energies in \eqref{Ezero}. Similarly, the $t^4$ term in $c_-$ 
 gives the $C_4=-(7/8)\zeta(3)$ coefficient as in \eqref{C4}.

Since the zero-point energies in \eqref{czero} match,
the following section proceeds to discuss how the expression \eqref{Zmuz} from \cite{Berg:2019jhh}
can be rewritten in a way that makes symmetries manifest.

\subsection{Modular covariance}
\label{modularcov}
The  {\it massive non-holomorphic
Kronecker-Eisenstein series} is defined as a double sum:
\begin{equation} \label{Edouble}
\mathcal E_{1, \mu}(z,\tau) = 2\sqrt{\mu \tau_2}\!\!\!\sum_{(c,d)\neq(0,0)}\!\!\! \frac{K_1 \left( 2 \pi \sqrt{\tfrac{\mu}{\tau_2}}  |c\tau + d |\right)}{|c \tau + d |} e^{2 \pi i ( c \beta-d\alpha)} \; , 
\end{equation}
where $K_1$ is a K-Bessel function, and  $ c_{\alpha,\mu}$  is given in \eqref{czero}.

The free energy is computed from minus the logarithm of the partition function $Z$ in eq.\  \eqref{Zmuz}.
Proposition 3.3 in \cite{Berg:2019jhh} is that \eqref{Edouble} is in fact equal to:
\begin{equation}  \label{E1first}
\mathcal E_{1,\mu}(z,\tau) =8 \pi c_{\alpha,\mu}\tau_2 +
 \sum_{n=-\infty}^{\infty} \sum_{\pm} \log \left(1- e^{-2\pi \tau_2 \sqrt{\frac{\mu}{\tau_2} + \left( n \pm \alpha\right)^2}+ 2 \pi i \left(n \pm \alpha\right)\tau_1\mp 2 \pi i \beta}\right) \; ,
\end{equation}
where  $\mu= m^2\tau_2$, and $z=\alpha\tau+\beta$  packages the two real quasi-periodicities
$\alpha$ and $\beta$ into a single complex number. The term $ c_{\alpha,\mu}\tau_2$  in \eqref{E1first} provides
the $c=0$ terms in \eqref{Edouble}.

To provide some basic intuition where \eqref{Edouble} comes from, first compare eq.\   \eqref{Edouble} to eq.\  \eqref{cBessel}, which is a one-dimensional lattice sum over $n$.
It is conceivable to {\it guess} the form of the two-dimensional lattice sum  \eqref{Edouble} over $c$ and $d$, recalling
$m=\sqrt{\mu/\tau_2}$. 

The special case $\alpha=\beta=0$ of \eqref{Edouble} 
for the free energy of a massive relativistic boson was derived in \cite{Ambjorn:1981xw}, eq.\ (3.6). 
In \cite{Kostov:2022pvy}, a similar double-sum representation is discussed as the path integral for a massive
theory with winding around the periods of the torus.
%For later use, note that in the representation \eqref{Edouble}, the mass parameter $\mu$ occurs only as a factor 
%multiplying the summation $|p\tau+q|$,
%whereas in \eqref{E1first}, the mass parameter $\mu$ is additive (``mixed in'') with the $n$ summation.
%

Now, the massless $\mu\rightarrow 0$ limit of the  double sum representation  \eqref{Edouble} recovers 
a basic example of an {\it automorphic form}: the ordinary non-holomorphic Kronecker-Eisenstein series
\be \label{classical}
E_{s}(z, \tau) =  \pi^{-s}\!\!\!\!\!\!\sum_{(c,d)\neq(0,0)}\! {\tau_2^s
\over |c\tau+d|^{2s} } e^{2\pi i (c\beta-d\alpha)}  \; . 
\ee
In this double sum form, it is easy to see that $E_{s}(0, \tau)$ is modular invariant for any $s$.
Interested readers may refer to the standard proof in Appendix \ref{class}, where I also try to explain why
it is natural to represent the fermion quasi-periodicities $\alpha$,$\beta$ by a marked point $z$ on the torus. 

A slightly more compact way to write eq.\ \eqref{Edouble} is as follows:
\be \label{Esimp}
\mathcal E_{1, \mu}(z,\tau) = 2\sqrt{\mu \tau_2}\sum_{\omega \neq 0}\frac{K_1 \left( 2 \pi \sqrt{\tfrac{\mu}{\tau_2}}  |\omega| \right)}{|\omega|} e^{2 \pi i \theta_{\omega}} \; , 
\ee
where $\omega=c\tau+d$ is the lattice vector  and the phase is $\theta_{\omega}=(c\beta-d\alpha)$,
and as before, $\alpha$ and $\beta$ take values 0 and 1/2 that distinguish the four fermion sectors.

In the product form \eqref{Zmuz}
 with a single product variable $n$, and  without expanding in the mass parameter, it takes a calculation 
 to prove modular invariance, and it seems to occur by accident. 
The following Mellin integral representation is given in \cite{Berg:2019jhh}:
 \begin{align} \label{invmellinZ}
\mathcal E_{1,\mu}(z,\tau) &=  {1 \over 2\pi i}\int_{c-i\infty}^{c+i\infty}  (\pi\mu)^{-s} \Gamma(s) E_{s+1}(z, \tau)ds \; ,
\end{align}
for any real number $c>0$. Since $\mu$ in \eqref{invmellinZ} is invariant by definition, and
the classical object \eqref{classical} is invariant, it follows that its massive deformation
\eqref{invmellinZ} is invariant. In other words, here modular transformation properties of the massive (deformed)
object follow from those of the corresponding massless (undeformed) object. I show \eqref{invmellinZ} below.

A more evocative way to see the invariance of \eqref{Edouble} is to construct it as a sum over images under the action of the modular group, a 
{\it Poincar\'e series}. This will not be pursued here, 
see \cite{Berg:2022feq} for details. 

To summarize this section: the well-known modular invariance of the ordinary non-holomorphic Eisenstein series $E_{s}(0, \tau)$ is a basic example of what ``manifest'' modular invariance means in this paper, and the undeformed partition function $Z_{\rm Ising}(0)$ can be expressed in terms of $E_{s}$. The point here is that also the deformed partition function $Z_{\rm Ising}(\mu)$ for $\mu \neq 0$ can  be expressed in  terms of the massive non-holomorphic Kronecker-Eisenstein series  ${\mathcal E}_{1,\mu}$ in \eqref{Edouble}, and its modular transformation properties follow from \eqref{invmellinZ}.

 \section{Zero-point energies: real integration}
 One can now view physical quantities in the deformed theory
 as non-holomorphic modular forms. This section concerns
only the cylinder limit $\tau_2\rightarrow \infty$,  where we 
 can only extract zero-point energies and neglect finite-size effects. Since $Z_{\rm Ising}\rightarrow |D|^2$ as $m\rightarrow 0$, and $Z_{\rm Ising}\rightarrow e^{-2\pi E_1}(1+\ldots)$
 as $\tau_2\rightarrow \infty$ ($q\rightarrow 0$),
 and in \eqref{czero}  we have $Z\rightarrow e^{-8\pi c_{\mu,\alpha}}(1+\ldots)$ as $\tau_2\rightarrow \infty$ ($q\rightarrow 0$), we
can extract $C_4$ from \eqref{czero} as (now writing $c_{1/2}(\mu)$ for $c_{1/2,\mu}$ to make the dependence more explicit when differentiating)
 \be
 C_4 = -2c_{1/2}(\mu)\Big|_{t^4} =-2\cdot {1\over 2}\tau_2^2{d^2 \over d\mu^2}\Big|_{\mu=0} c_{1/2}(\mu) 
 =-\tau_2^2{d^2 \over d\mu^2}\Big|_{\mu=0} c_{1/2}(\mu) 
 \ee
 where the $\tau_2^2$ comes from $m^2=\mu/\tau_2$. By the arguments in the previous secion, it could be a cause of concern why a physical quantity 
 should depend on $\tau_2$, that depends on the choice of fundamental domain (gauge). The situation will be clearer
 when we consider the full (two-dimensional) expansion of the free energy below.
For $\alpha=1/2$ in \eqref{czero}, and slightly generalizing the integrand with factors of $x^{s-1}$ 
and $1/(2\pi)^s$ that reduce to those in \eqref{czero} when $s=1$,
we have
\bea
 {d^2 \over d\mu^2}\Big|_{\mu=0}c_{1/2,{\mu}} &=&\frac1{(2\pi)^{2s}}  \int_0^\infty \!\!x^{s-1} \sum_{\ell \geq 1} \cos( 2\pi \ell/2)e^{-\ell^2x} {d^2 \over d\mu^2}\Big|_{\mu=0}e^{-\frac{\pi^2\mu/\tau_2}{x}}dx \nonumber \\
&=& \frac1{(2\pi)^{2s}}\int_0^\infty x^{s-1}{1 \over 2}(\vartheta_4(0,ix/\pi)-1)\frac{\pi^4}{\tau_2^2 x^2} \, dx  \; . \label{first}
\eea
Note that the $x^{-2}$ from the differentiation with respect to $\mu$ lowered the power of $x$ 
by 2. Here $\vartheta_4$ is a  Jacobi theta function
\be
\vartheta_4(0,\tau) = \sum_{n=-\infty}^{\infty} (-1)^n e^{\pi i n^2 \tau} \; . 
\ee
The integral \eqref{first} is a Mellin transform of a Jacobi theta function. A similar integral
was computed already by Riemann in his 1859 paper on analytic number theory (see e.g.\ \cite{Edwards}, or Problem 6.15 in \cite{Apostol}), and a
version of the integral in  \eqref{first} is given for example in
 \cite{Bateman} section 6.9:
\be \label{integral}
\int_0^\infty \!x^{\tilde{s}-1}(\vartheta_4(0,i x^2)-1) dx = (2^{1-\tilde{s}}-1)\pi^{-\tilde{s}/2}\Gamma(\tilde{s}/2)\zeta(\tilde{s}) \; . 
\ee
To use this in \eqref{first}, change variable of integration to $y^2=x/\pi$:
\be
\frac1{(2\pi)^{2s}}\int_0^\infty x^{s-1}{1 \over 2}(\vartheta_4(0,ix/\pi)-1)\frac{\pi^4}{\tau_2^2 x^2} \, dx  \\
= 2^{-2s}\pi^{2-s} \tau_2^{-2}\int_0^\infty y^{2s-5}(\vartheta_4(0,iy^2)-1) \, dy  \label{Mellt} \;. 
\ee
We identify $\tilde{s}-1$ in \eqref{integral} with $2s-5$ in \eqref{Mellt}, so $\tilde{s} = -2$ for $s=1$. Now the right-hand side of \eqref{integral}  seems divergent
for $\tilde{s} = -2$, but the functional equation for the
Riemann zeta function $\zeta(s)$ states that the right-hand
side of \eqref{integral} can be reflected $\tilde{s}\rightarrow 1-\tilde{s}$ as follows:
\be \label{reflect}
\pi^{-\tilde{s}/2}\Gamma(\tilde{s}/2)\zeta(\tilde{s}) = \pi^{-1/2+\tilde{s}/2}\Gamma(1/2-\tilde{s}/2)\zeta(1-\tilde{s})\; . 
\ee
Using \eqref{reflect} and \eqref{integral} in \eqref{Mellt}, and setting $s=1$ ($\tilde{s} = -2$), we find
 \be
 C_4 = -2\tau_2^2{d^2 \over d\mu^2}\Big|_{\mu=0} c_{1/2}(\mu)  =
 -\tau_2^2 \cdot 2^{-2}\pi\tau_2^{-2}\cdot (2^3-1)\pi^{-3/2}\Gamma(3/2)\zeta(3) =  -{7 \over 8}\zeta(3) \; .
 \ee
This reproduces $C_4$ of earlier sections. 

The reader not familiar with this type of calculation may find this section more complicated than the original
calculations in earlier sections. In particular, here we needed $s$-regularization, and we needed the integral \eqref{integral}.
As already noted, the $\tau_2$ in intermediate steps may also seem inconvenient. 
The purpose of this section is not to be convenient, but to show using elementary methods that
the possibly unfamiliar \eqref{Edouble} gives the same results as found earlier. The next section describes
a more efficient alternative calculation, by residue calculus.
 
\section{Zero-point energies: contour integration}
\begin{figure}[h]
\begin{center}
\includegraphics[width=0.45\textwidth]{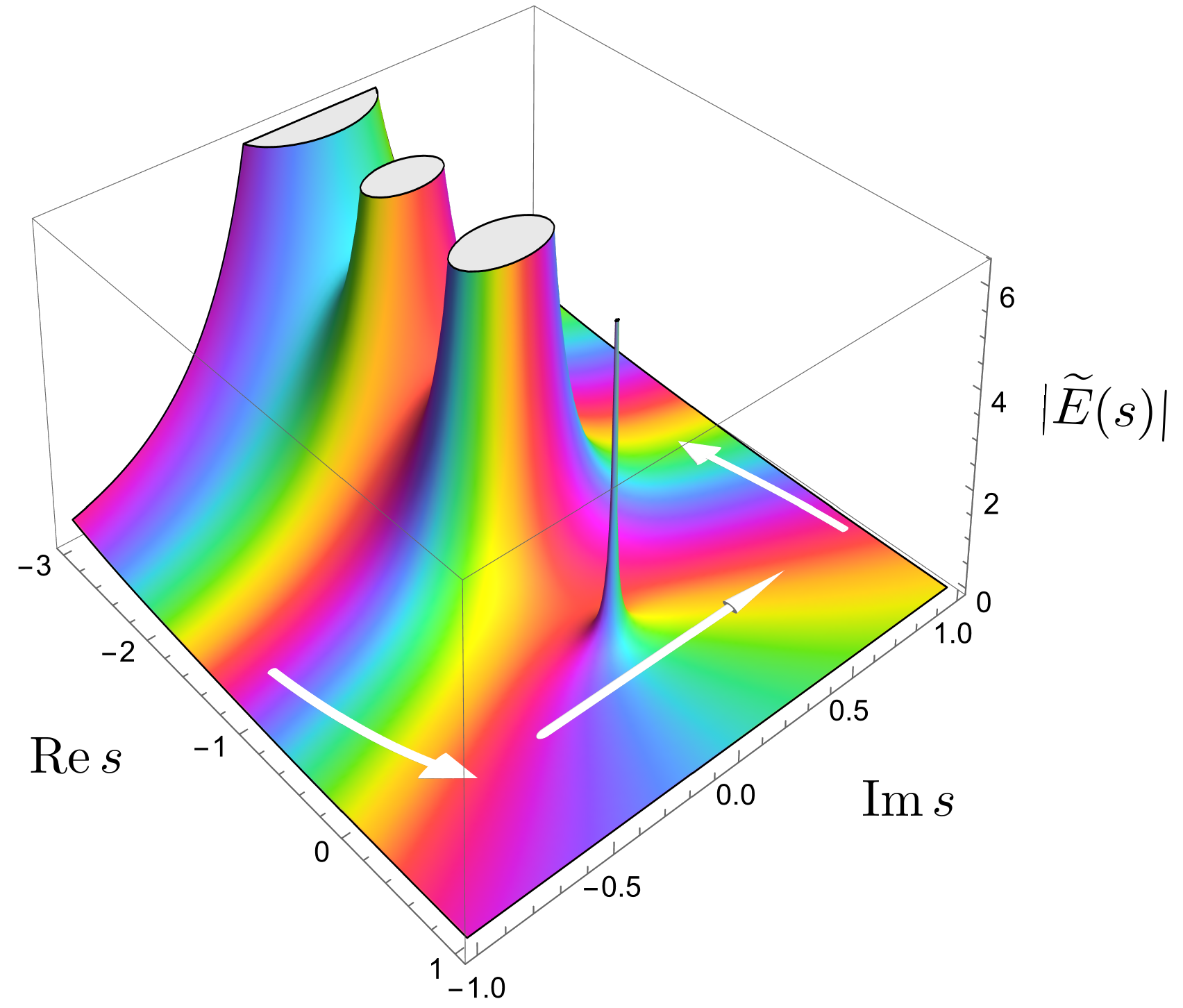}
\caption{The Mellin-dual energy $\widetilde{E}(s)$ in the complex $s$ plane. The three arrows give an idea of the integration contour,
which closes at infinity. The  argument ${\rm arg}(\widetilde{E}(s))$ sets the color.}
\label{fig:contour}
\end{center}
\end{figure}

\begin{figure}[h]
\begin{center}
\includegraphics[width=0.5\textwidth]{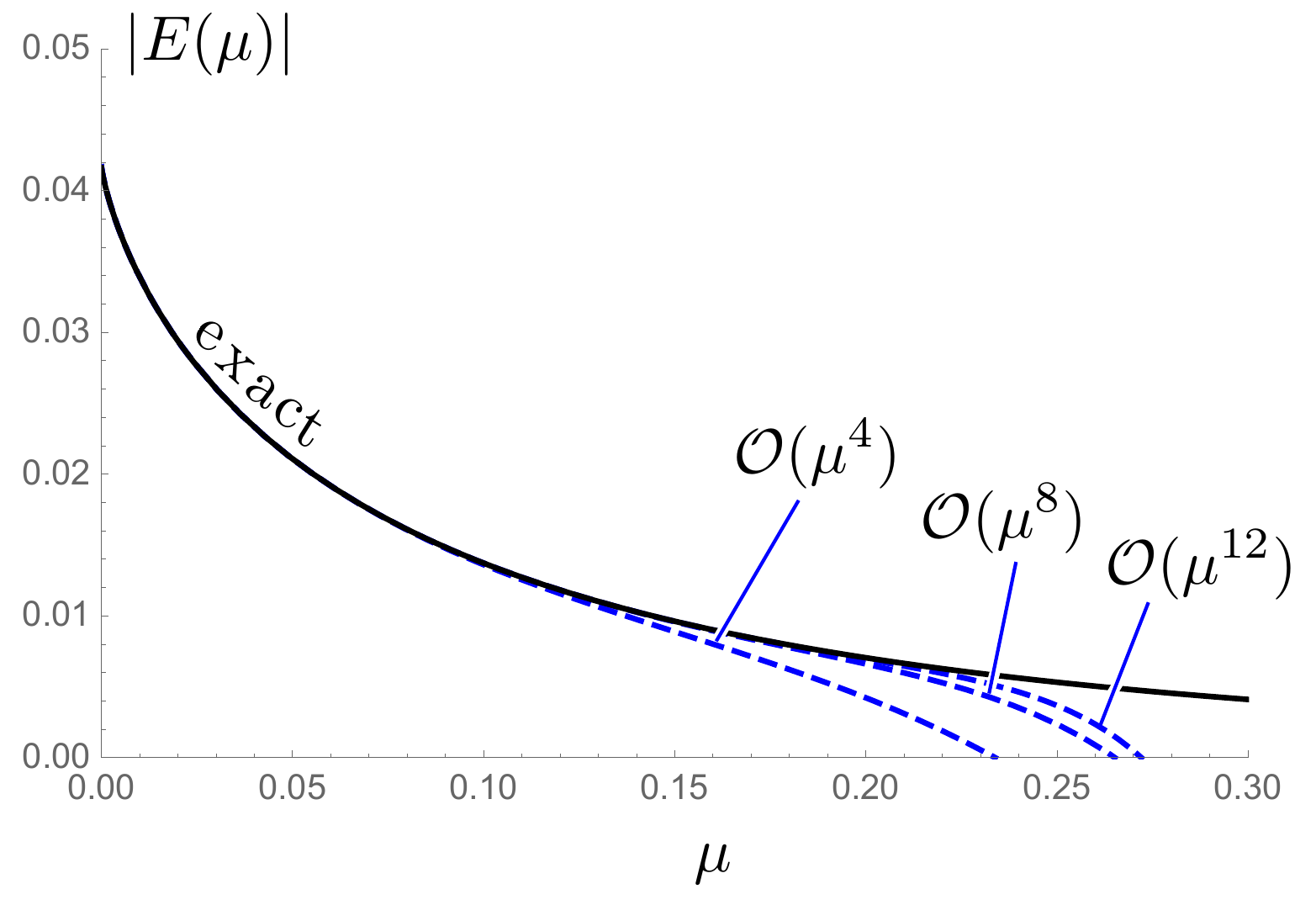}
\caption{{\small Black: Bessel sum representation of the zero-point energy $E(\mu)$ in \eqref{restot}. Blue dashed: truncation of eq.\ \eqref{close} at order 4,8,12 in $\mu$.
For large mass, $E(\mu)$ becomes negligible, as expected.}}
\label{fig:Emu}
\end{center}
\end{figure}

Here is an alternative to the calculation in the previous section, that does not require
integrating Jacobi theta functions, and gives some additional insight.
Equation \eqref{invmellinZ} indicates that it may be useful to Mellin transform the zero-point energy $E_1(\mu) = 2c_{1/2,\mu}$
\be
E(\mu) =2{\sqrt{\mu/\tau_2} \over 2\pi}\sum_{\ell=1}^{\infty} (-1)^l  {K_1(2\pi \ell \sqrt{\mu/\tau_2}) \over \ell} \; , 
\ee
where $\sqrt{\mu/\tau_2}=m$, and $E$ is used as a shorthand for $E_1$ in this section. 
Mellin transforming from $\mu$ to $s$ gives
\bea \label{MellinE}
\widetilde{E}(s) = \int_0^{\infty} \mu^{s-1}E(\mu) d\mu &=& -{1 \over 2}\pi^{-2s-2}\tau_2^s \, s\,  \Gamma(s)^2\sum_{\ell=1}^{\infty}(-1)^{\ell+1} \ell^{-2s-2}\\
&=& -(2^{2s+1}-1)(2\pi)^{-2s-2}s\tau_2^s \zeta(2s+2)\Gamma(s)^2   \; . 
\eea
The Mellin transform $\widetilde{E}(s)$ has poles
at zero and negative integers, and falls off at complex infinity. The Mellin contour 
for the transformation back (effectively a Mellin-Barnes representation)
can be chosen to be the vertical line $s=1/2+ix$ for real $x$, which 
can be closed
by a semicircle at infinity on the left side of the $s$ complex plane, as indicated in figure \ref{fig:contour}. 
The contour integration picks up residues as follows
(in the remainder of this section $\tau_2=1$ to reduce clutter, it can be restored by $\mu\rightarrow \mu/\tau_2^2$):
\bea
E(\mu) &=& \sum_{n=0}^{\infty} {\rm Res}[\widetilde{E}(s) \mu^{-s},s=-n]  \label{restot} \\
&=&-{1 \over 24} -c_1\mu - { 7\over 2}\pi^2\zeta'(-2)\mu^2-{31 \over 12}\pi^4 \zeta'(-4)\mu^3
-{127 \over 144}\pi^6\zeta'(-6)\mu^4+\ldots \label{residue1} \\
&=&-{1 \over 24} -c_1\mu +{7 \over 8}\zeta(3)\mu^2 -{31 \over 16} \zeta(5)\mu^3 +{635 \over 128}\zeta(7)\mu^4
-{3577 \over 256}\zeta(9)\mu^5+{42987 \over 1024}\zeta(11)\mu^6+\ldots \label{goodexp} \\
&=& -{1 \over 24} -c_1\mu   +1.0518\mu^2 -2.00905 \mu^3
+5.00236\mu^4 - 14.0007 \mu^5 +42.0002\mu^6 + \ldots  \label{close}
\eea
where in the third equality
\be
\zeta'(-2n) = (-1)^n {(2n)! \over 2(2\pi)^{2n}}\zeta(2n+1)
\ee
which follows from \eqref{reflect}. 
Also, in  \eqref{residue1}, the coefficient of the linear term is $c_1 = (2\gamma-1+\log 4\mu)/4$, and $\gamma$ is the Euler-Mascheroni constant. The 
$E(\mu)$ above is plotted in  fig.\ \ref{fig:Emu}.
The coefficients
in \eqref{close} become closer 
and closer to integers. The absolute values of those limiting integers are the Catalan numbers:
\be \label{Catalan}
C_n = {(2n)! \over (n+1)!\, n!} = 1,2,5,14,42,132,429, \ldots  \quad n=1,2,\ldots
\ee
At first glance the appearance of near-integer coefficients seems a little surprising, since
an infinite number of odd zeta values are irrational\cite{Rivoal}.
But there are many approximations for $\zeta(2n+1)$ in terms of more elementary expressions, for example by Ramanujan  \cite{Berndt},
and eq.\ \eqref{close} above displays a very simple such  leading approximation, by rational numbers:
\be \label{zetaapprox}
\zeta(2n+1)\approx {1 \over 1-2^{1-2n}}  = {8 \over 7}, \, {32 \over 31} , \,{128 \over 127}, \,{512 \over 511},\,{2048 \over 2047}  \ldots \qquad n=1,2,\ldots
\ee
For example, the exact value $\zeta(11)\approx 1.000494$ is approximated by $2048/2047=1.000489\ldots$. It is easy to see why this is a decent
approximation: by definition $\zeta(11) = 1+1/2^{11} + \ldots$, so if we truncate to the first nontrivial term, $\zeta(11)\approx 1+1/2^{11} = 2049/2048$. The ratio $2048/2047$ in \eqref{zetaapprox} is just a shift by one in numerator and denominator, that slightly
increases the value, and brings it slightly closer to the exact value.

The generating function for (alternating) Catalan numbers \eqref{Catalan}   gives an asymptotic, ``resummed'' approximation to the higher-order terms in the energy:
\be  \label{generate}
E(\mu)-\left(-{1 \over 24} -c_1\mu\right) \approx  \sum_{n=1}^\infty (-1)^{n+1} C_n \mu^{n+1} = {1 \over 2}+\mu - \sqrt{1+4\mu} \; . 
\ee
For large $\mu$, a fractional power of $\mu$ appears, which was not so obvious from the expansion. The Bessel representation clarifies this somewhat, since it also behaves as a fractional power
at infinity. 

Another comment is that since  the coefficient $C_4=-(7/8)\zeta(3)$ 
appears as a single term in  \eqref{goodexp}, specifically from  the pole at $s=-2$. Therefore, the residue
calculation \eqref{res}
can be viewed as an indirect proof of the integration formula \eqref{integral} for the Jacobi theta function. 

Now that the smoke has cleared,  we see that when we power expand the Bessel function,  the sums naively diverge and should be zeta-function regularized. The point in this paper is that away from criticality, there is no reason to ruin the exponential convergence due to the Bessel function by series-expanding it in the deviation from the critical point, once we are convinced that we can achieve manifest modular invariance without such expansion. 

Finally,  we can see the relation to the $\lambda$ integral representation
in earlier section, and eq.\ \eqref{generalexp}, which I repeat here for convenience:
\be
\sum_{n=2}^{\infty}
(-1)^{n+1}(2^{2n}-2) {\Gamma(n-{\scriptstyle {1 \over 2}}) \over 4\sqrt{\pi} n!}
\zeta(2n-1)t^{2n} = -{7 \over 8}\zeta(3)t^4 + {31 \over 16}\zeta(5)t^6
-{635 \over 128} \zeta(7) t^8 + {3577 \over 256 }\zeta(9)t^{10}+ \ldots
\ee
in agreement with eq.\ \eqref{close}. 

Again, these comparisons are made to connect to existing literature, that typically considers the
 $\mu\rightarrow 0$ limit, whereas for nonzero $\mu$, using the Bessel function representation is preferable
to series expansion, since the former converges exponentially. In the next section, this will be extended to the full (finite-size) torus sum, not just the zero-point energies in the $\tau_2\rightarrow \infty$ limit. 

\section{Invariant mass expansion}
\label{2D}
Here we reproduce the $\mu$ expansion \eqref{expansion}, as the 2D analog of the 1D $\mu$ expansion above.
The partition function $Z_{\rm Ising}(\mu)$ is written in terms of the free energies of each fermion sector,
here the massive Kronecker-Eisenstein series \eqref{Edouble}, as
\be \label{Zstart}
Z_{\rm Ising}(\mu) = \sum_{i=1,2,3,4} e^{-{\mathcal E}_{1,s}(z_i,\tau)}
\ee
where $i=1,2,3,4$ corresponds to\footnote{For comparison with  \cite{Saleur1987}, recall from earlier
sections that their convention $(1/2,0) \rightarrow (0,1/2)$ here.}  $ z = \alpha + \beta \tau$ for $(\alpha,\beta) = (1/2,1/2),(1/2,0),(0,0),(0,1/2)$. 
(See Appendix \ref{invmu} for a comment on the periodic-periodic sector.)

From \eqref{Zstart} we can write down
the expansion of minus the total free energy, $\ln Z_{\rm Ising}(\mu)$, as a weighted sum of derivatives of Eisenstein series evaluated at the critical point $\mu=0$:
\be
\ln Z_{\rm Ising}(\mu) = \ln Z_{\rm Ising}(0) - \mu {\sum_i Z_i {\mathcal E}'(z_i) \over Z_{\rm Ising}(0)} 
-{\mu^2\over 2} \left(  {\sum_i Z_i ({\mathcal E}''(z_i)-{\mathcal E}'(z_i)^2) \over Z_{\rm Ising}(0)^2} +{(\sum_i Z_i {\mathcal E}'(z_i))^2 \over Z_{\rm Ising}(0) }   \right)+\ldots  \label{muexp0}
\ee
where the  $i$ sum only has three terms, due to the periodic-periodic sector dropping out by its partition function vanishing
at the critical point. 
The ``building blocks'' here are massive Kronecker-Eisenstein series \eqref{Edouble}, with the shorthand
${\mathcal E}(z)={\mathcal E}_{1,\mu}(z,\tau)$ and their derivatives with respect to the invariant $\mu$, all evaluated at the massless point $\mu=0$. It follows that \eqref{muexp0} is modular invariant if continued to any order, without further manipulation as was necessary in  \eqref{expansion}. 

Following the same logic as in the previous section, more important than the $\mu$ expansion in \eqref{muexp0} is the simple statement is that $Z_{\rm Ising}(\mu)$ is itself proven to be modular invariant nonperturbatively in $\mu$. Series expansion like \eqref{muexp0}  is only useful to connect quantities in the noncritical theory to those in the critical theory.

To make this connection, we would like explicit expressions for ${\mathcal E}'(z)$ and ${\mathcal E}''(z)$ at $\mu=0$,
that are generated by series-expanding ${\mathcal E}$ in $\mu$. First a demonstration of eq.\ \eqref{invmellinZ} as promised, except it is easier to show the forward transformation:
\bea \label{FEB}
 \int_0^{\infty} d\mu\; \mu^{s-1}{\mathcal E}_{1,\mu}(z,\tau)  &=&\int_0^{\infty}  d\mu\;  \mu^{s-1}  2\sqrt{\mu \tau_2}\sum_{\omega}\frac{K_1 \left( 2 \pi \sqrt{\tfrac{\mu}{\tau_2}} | \omega| \right)}{|\omega|} e^{2 \pi i \theta_{\omega}} \nonumber \\
 &=&  \pi^{-2s-1}s\Gamma(s)^2 \sum_{\omega}{\tau_2^{s+1} \over |\omega|^{2s+2}}e^{2 \pi i \theta_{\omega}}
 =\pi^{-s}\Gamma(s)\Gamma(s+1) E_{s+1}(z,\tau) \; . 
\eea
Similarly to the 1D case above, inverting the Mellin transformation gives a sum of residues from the Mellin contour integration:
\bea \label{muexp}
{\mathcal E}_{1,\mu}(z,\tau)  &=& \sum_{n=0}^{\infty} {\rm Res}[\pi^{-s}\Gamma(s)\Gamma(s+1) E_{s+1}(z,\tau) \mu^{-s},s=-n]  \label{res} \\[-1mm]
&=&  E_1(z,\tau) -\pi \mu (E_0'(z,\tau)+\log(\pi \mu)+2\gamma-1) -{1 \over 2}{\pi^2 \mu^2}E_{-1}'(z,\tau) -{1 \over 12}{\pi^3 \mu^3}E_{-2}'(z,\tau)+ \ldots \nonumber \\
&=&  E_1(z,\tau) -\pi \mu( \widetilde{E}_1^{\rm reg}(z,\tau)+\log(\pi \mu)-1) -{1 \over 2}{\pi^2 \mu^2}\widetilde{E}_{2}(z,\tau) +{1 \over 3}{\pi^3 \mu^3}\widetilde{E}_{3}(z,\tau)+ \ldots   \nonumber 
\eea
using from appendix \ref{class} the reflection formula (functional relation) eq.\ \eqref{refl} and its consequence \eqref{E0vals}.
The tilde is needed to keep track of where the shift is: in the phase (original version, with a pole at $s=0$), or in the denominator (reflected version with a tilde, with a pole at $s=1$, in \eqref{Edual}).
 In fact, the reflected (tilded) version eq.\  \eqref{Edual} in appendix \ref{class} is where  Itzykson and Saleur  \cite{Saleur1987} started from, since the denominator is
the eigenvalue for quasiperiodic fields.

In \eqref{muexp}, the vanishing of the Eisenstein $E_{n}$ for negative integer $n$ was used. This is simpler than $\zeta(n)$ in the 1D case above, that only vanishes for  {\it even} negative integers.

From \eqref{muexp}, we can read off the  ${\mathcal E}'$ and ${\mathcal E}''$  terms needed for the $\mu$ expansion of the free energy away from criticality in \eqref{muexp}, the analogues of the $\mu$ (or $t^2$) and $\mu^2$ (or $t^4$) terms in the 1D case:
\be \label{derivatives}
{\mathcal E}'(z) =  -\pi (\widetilde{E}_{1}^{\rm reg}(z)+\log(\pi \mu)-1) \; , \quad 
{\mathcal E}''(z) = -\pi^2 \widetilde{E}_{2}(z) \; .
\ee
The non-curly $E_s$ are ordinary Kronecker-Eisenstein  sums, that are Jacobi group covariant and convergent, the sole exception in general being  $\widetilde{E}_{1}(z)$, that is divergent if evaluated by double sums. With only zeta-regularization, this divergence would have appeared as a pole at $s=1$, that would need to be subtracted by hand. Here
with mass,  the Mellin contour by definition avoids the pole at $s=1$, so the threat of a divergence from ${\mathcal E}'$ in \eqref{derivatives} is removed by construction, and replaced by a $\log \mu$ term. Note that as far as eq.\ \eqref{muexp} is concerned,
this $\log \mu$ comes with a factor $\mu$ in front, so it does not make the free energy itself divergent, it provides the expected singularity of the specific heat. 

The remaining $\widetilde{E}_{1}^{\rm reg}(z)$, while regular, can be inconvenient for numerical computations, as discussed further around eq.\ \eqref{E0vals} in the appendix.
If instead of  the partition function we wanted to compute correlation functions in the noncriticial Ising model, $\widetilde{E}_{1}^{\rm reg}(z)$ does not appear. This is because
if we allow $E_s(z,w)$ to have quasiperiodicities in both original and reflected (tilded) versions,  there is no pole either at $s=0$ or $s=1$, as discussed in  \cite{Berg:2019jhh}.

\begin{figure}[h]
\begin{center}
\includegraphics[width=0.5\textwidth]{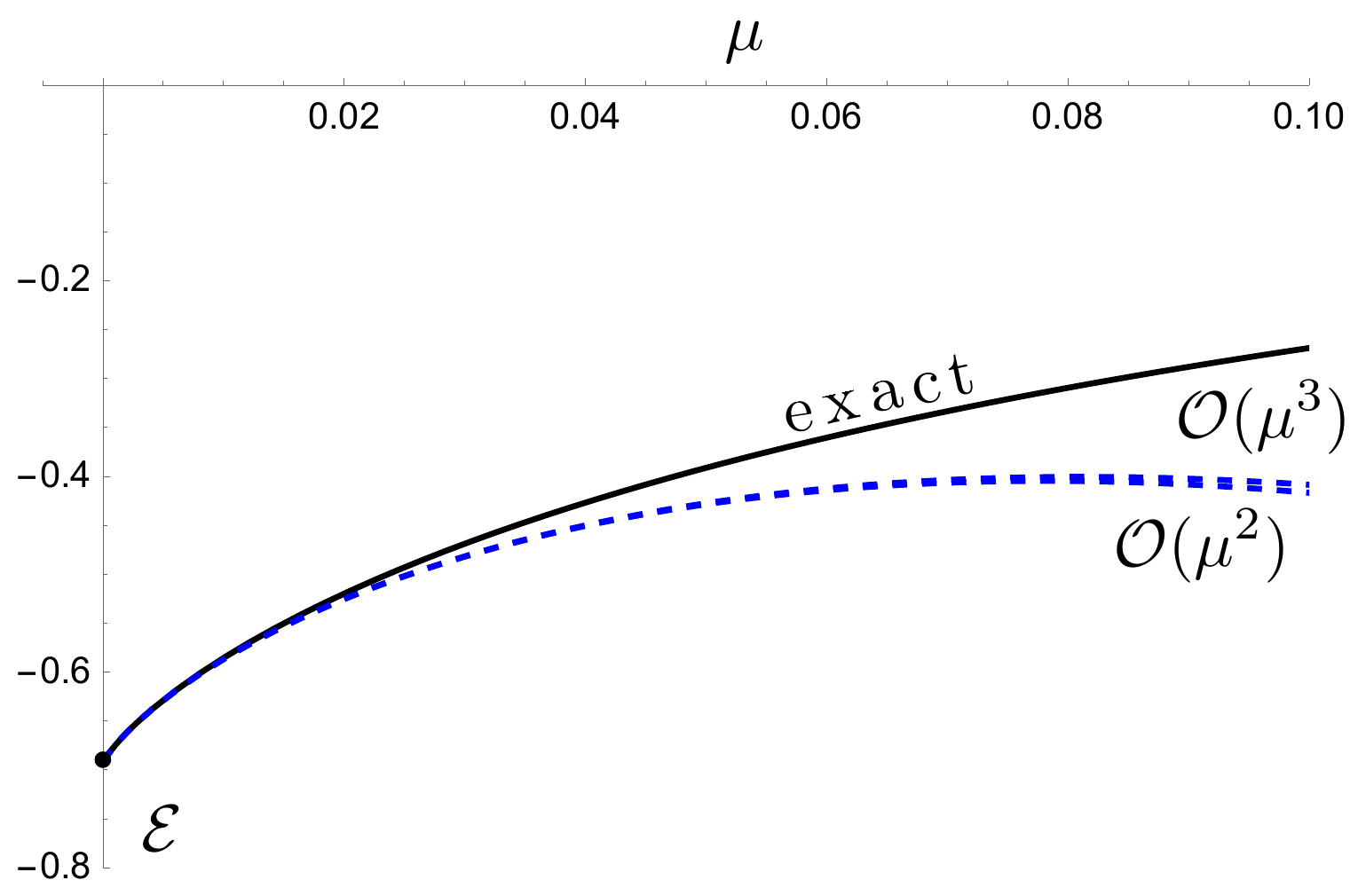}
\caption{{\small Black: the free energy building block ${\mathcal E}_{1,\mu}$ in \eqref{FEB}. Blue dashed: truncation of eq.\ \eqref{muexp} at order 2,3 in $\mu$.
The $E_0'$ value was fixed to $-0.1$ to match the exact result (cf.\ eq.\ \eqref{E0vals}).}}
\label{figE1mu}
\end{center}
\end{figure}

To summarize, the new result in this section is the mass expansion eq.\  \eqref{muexp0} together with eq.\ \eqref{muexp},  that is manifestly modular invariant,
and can be written down to any desired order,  where eq.\ \eqref{derivatives} is sufficient up to and including order $\mu^2$. A plot of  eq.\  \eqref{muexp0} 
is provided in fig.\ \ref{figE1mu}. 

A final comment: if we are only interested in derivatives, we can also use that the massive Kronecker-Eisenstein series
${\mathcal E}$ satisfies a partial differential equation of the schematic form
\be
\mu^2 {\partial^2 \over \partial \mu^2}{\mathcal E} = \Delta_\tau {\mathcal E} \; , 
\ee
which can be simple to compute.
This direction will not be pursued further here, but it is natural from the point of view of the mathematics ({\it Jacobi-Maass forms})
to be able to reformulate differentiation with respect to the external parameter $\mu$ in terms of the intrinsic 
Laplace operator  $\Delta_\tau$. In particular, this means that the theory at criticality ``knows'' about at least some properties away from criticality.

\section{Finite-size effects}
\label{finite}
Ferdinand and Fisher \cite{Ferdinand:1969zz} viewed the $\tau_2\rightarrow \infty$ limit of the torus as the cylinder Onsager model,
and finite $\tau_2$ as finite-size effects. These finite-size effects should also be accessible with the methods discussed in this paper.

In \cite{Ivashkevich} (and reviewed in \cite{Izmailian:2017vtg}, see also \cite{Salas:2000wq}), the Ferdinand-Fisher free energy was generalized as
\be \label{Iva}
F_{T = T_{\rm c}}(\rho,S)=f_{\rm bulk}(\rho)A
+f_0(\rho) +f_1(\rho)A^{-1} + \ldots
\ee
for $\rho$ the aspect ratio of the torus (assuming that $\tau_1=0$), and $A$ the torus area (which was called $S$ in\cite{Ivashkevich}),
with  the expansion coefficients
\be
f_{\rm bulk}(\rho) =-\ln\sqrt{2}-2{\gamma \over \pi} \; , \quad f_0(\rho) = -\ln {\vartheta_2(i \rho)+\vartheta_3(i\rho)+\vartheta_4(i\rho)
\over 2\eta(i\rho)} \; . 
\ee
Similarly,
the specific heat
\be
c_{T = T_{\rm c}}(\rho,A)=c_{\rm bulk}(\rho,A)A
+c_0(\rho) +c_1(\rho)A^{-1} + \ldots
\ee
with the expansion coefficients
%\footnote{Ferdinand-Fisher eq.\ (4.21), apart from normalization seems to agree
%with this $c_{\rm bulk}(\rho)$. } 
(now with the argument $i\rho$ suppressed),
\be \label{cbulk}
c_{\rm bulk}(\rho,A)
={8 \over \pi}\left(\ln\sqrt{A \over \rho}
+\ln {2^{5/2} \over \pi}+\gamma-{\pi \over 4}\right) -4\rho\left({2\eta \over \vartheta_2+\vartheta_3+\vartheta_4}\right)^2-{16 \over \pi}{\vartheta_2 \ln \vartheta_2 + \vartheta_3\ln \vartheta_3+\vartheta_4 \ln \vartheta_4
\over \vartheta_2+\vartheta_3+\vartheta_4}
\ee
with $\gamma$ the Euler-Mascheroni constant,
and
\be
c_0(\rho) = -2\sqrt{2}\sqrt{\rho}{2\eta
\over \vartheta_2+\vartheta_3+\vartheta_4} \; . 
\ee
Here we can view the area as set by $m$, and the aspect ratio
$\rho$ is set by $\tau_2$,
but we want to view each coefficient
as a function of $\tau$, not just $\tau_2$. We can interpolate
between square lattice $\tau=i\tau_2$ and triangular lattice,
$\tau=\rho e^{\pi i /3}$. 

As a first comparison, note that the Eisenstein series is the logarithm of the $\vartheta$ functions in the partition function,
so the last term in $c_{\rm bulk}$ in eq.\ \eqref{cbulk} corresponds to $\sum Z_i {\mathcal E}'(z_i)$. 

Each term in the expansions of $F$ and $c$ here are modular forms, rather than invariants. So again,
although \eqref{Iva} from \cite{Ivashkevich}  has advantages compared to \cite {Saleur1987},  each term still needs to be checked separately,
and the coefficients change if we change the lattice vectors (frame for modular transformations). 
In the $\mu$ expansion above, each building block is Jacobi group covariant, so the terms in the sum merely permute under S and T, and the coefficient
at each order in $\mu$ is separately invariant. 

On the other hand, for numerical purposes, representations in terms of Jacobi theta functions can be useful for their rapid convergence,
compared to the massless Eisenstein double-sum series of eq.\ \eqref{classical} and appendix \ref{class}. 
However,  the massive Kronecker-Eisenstein
series also converge exponentially.

\subsection{Modular transformations and numerics}
\begin{figure}[h]
\begin{center}
\includegraphics[width=0.5\textwidth]{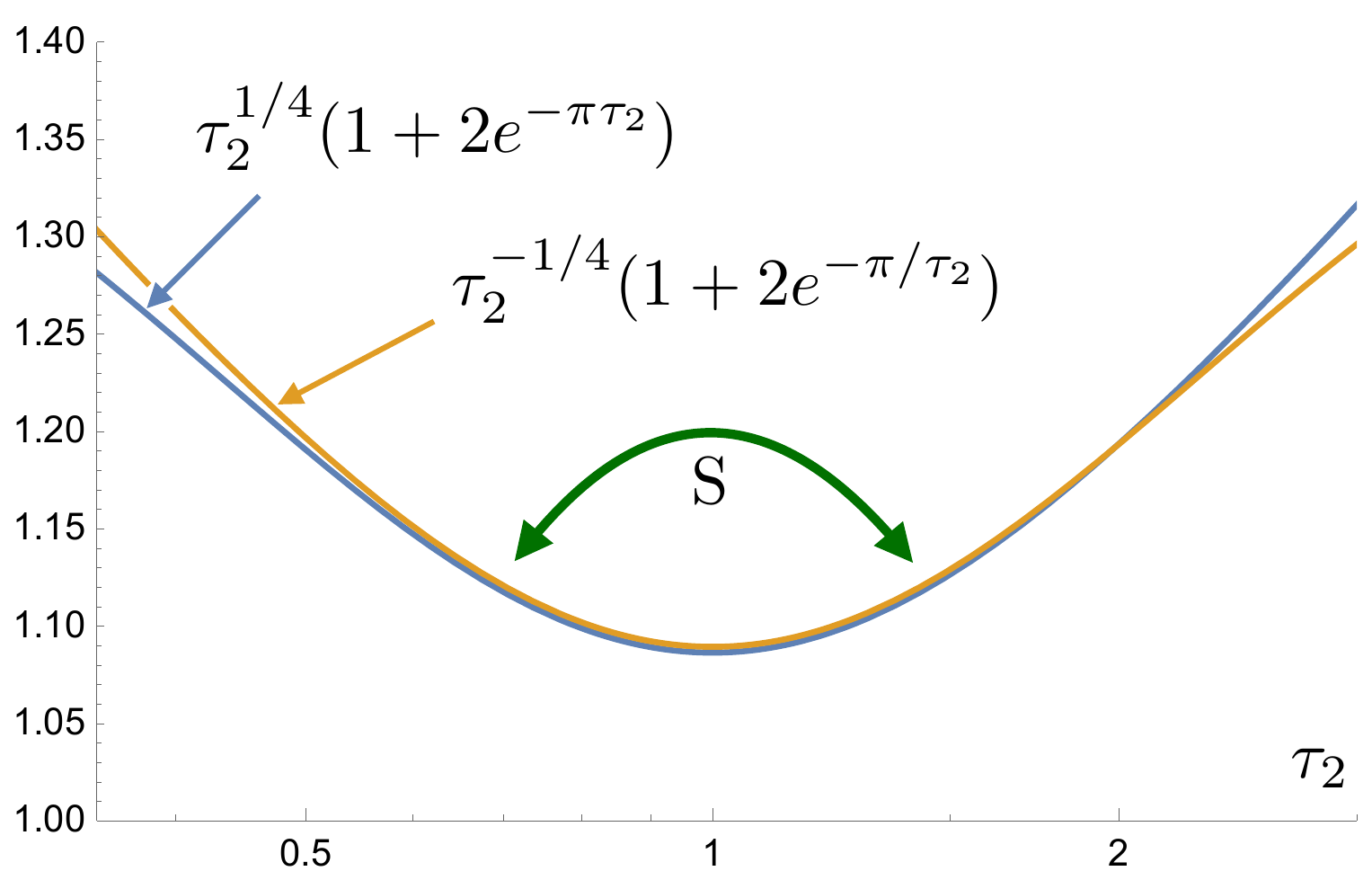}
\caption{Plotting two truncated expressions with logarithmic scale on the horizontal axis. The untruncated 
$|\tau_2^{1/4}\vartheta_3(i\tau_2)| = |\tau_2^{-1/4}\vartheta_3(i/\tau_2)| $ would have been strictly symmetric under S.}
\label{fig:thetaS}
\end{center}
\end{figure}

Here is a toy illustration of the difference between ordinary Jacobi theta series, massless Kronecker-Eisenstein series, and massive Kronecker-Eisenstein series,
for the purpose of numerics.
Consider the Jacobi $\vartheta_3$ function, that 
for a square torus lattice $\tau=i\tau_2$ satisfies 
the S transformation property
\be \label{thetaS}
\tau_2^{1/4}\vartheta_3(i\tau_2) = \tau_2^{-1/4}\vartheta_3(i/\tau_2)  \; . 
\ee
(The factors of $\tau_2^{1/4}$ in \eqref{thetaS} are such that the two sides of this equation are separately invariant.) If we plot this relation as a function of $\tau_2$
with a logarithmic scale on the horizontal axis, the plot should be symmetric under reflection around $\tau_2=1$, since $\log (1/\tau_2)=-\log \tau_2$. 
In fig.\ \ref{fig:thetaS}, the two sides of   \eqref{thetaS} are truncated to the first nontrival term and plotted.
This is just a toy illustration, since a plot of \eqref{thetaS} with just a few more terms included in the trunction
 looks more symmetric. In physical applications, more complicated expressions may
take more terms to make the symmetry apparent in the plot.

For comparison, consider  $E_1(z_3,\tau_2)$ where $z_3 = 1/2+\tau/2$ and $\tau=i\tau_2$, that is S-invariant: $E_1(z_3,i\tau_2)=E_1(\tilde{z}_3,i/\tau_2)$
where $\tilde{z}_3=z_3/\tau = 1/2-i/(2\tau_2)$. Here invariance holds even at  the first nontrivial truncation. This  can be shown directly from the general expression in appendix \ref{class},
but perhaps a simple example is of some use.

Truncated to $|c|,|d|\leq 1$, the double sums give
\be \label{E1S}
{\pi \over 2} E_1(z_3,i\tau_2)\Big|_{|c|,|d|\leq 1} = -{1+\tau_2^4 \over \tau_2+\tau_2^3} \; , \qquad
{\pi \over 2} E_1(\tilde{z}_3,i/\tau_2)\Big|_{|c|,|d|\leq 1} = -{1+\tau_2^4 \over \tau_2+\tau_2^3} \; ,
\ee
so for this  first nontrivial truncation, the S-related expressions in \eqref{E1S} are manifestly the same, we do not need to plot them. However, the double sum converges slowly (in fact conditionally,
since $s=1$ in $E_s$). So although symmetry is manifest at each order of truncation, the plots themselves may change significantly
when the numbers of terms included is increased. 

Now, a property like $E_1(z_3,i\tau_2)=E_1(\tilde{z}_3,i/\tau_2)$ above extends to the massive Kronecker-Eisenstein series ${\mathcal E}_{s,\mu}$ when truncated:
\be \label{curlyEs}
{1 \over 4\sqrt{\mu}} {\mathcal E}_{1,\mu}(z_3,\tau_2)\Big|_{|c|,|d|\leq 1} ={1 \over \sqrt{\tau_2}}K_1(2\pi\sqrt{\mu\tau_2}) + \sqrt{\tau_2} K_1(2\pi\sqrt{\mu/\tau_2})
+{2 \over \sqrt{\tau_2+1/\tau_2}}K_1(2\pi \sqrt{\mu(\tau_2+1/\tau_2)}) \; . 
\ee
Indeed, $ {\mathcal E}_{1,\mu}(\tilde{z}_3,i/\tau_2)$ truncated to $|c|,|d|\leq 1$ gives the same expression as in \eqref{curlyEs}: the first two terms switch place under $\tau_2\leftrightarrow 1/\tau_2$,
and the last term stays the same. Unlike for the $\mu=0$ case, plots of ${\mathcal E}_{1,\mu}$ for $\mu>0$ will not change much when
 including more terms in the truncation, since the Bessel K function
decreases exponentially with increasing argument. 

Perhaps this order-by-order invariance together
with the relatively quick convergence  can be of some use for numerics, beyond this toy example. 

 \section{Outlook: physical quantities as modular invariants}
One very simple point of this paper is that the parameter $m$ in the differential operator $\partial\bar{\partial} + m^2$ transforms
under modular transformations, but we can construct an invariant mass parameter by simply multiplying by $\tau_2$, 
to make the invariant $\mu=m^2\tau_2$. 
In the example of running central charge, this does not matter much,
 since by taking the limit $\tau_2\rightarrow \infty$,
we have already picked a {\it modular frame}, i.e.\ we can no longer perform an S transformation that inverts $\tau_2$,
and only the zero modes in the ``horizontal'' $\omega_1$ direction contribute,
so the problem is effectively one-dimensional. 
Section \ref{2D} gives an example what can be done in a truly two-dimensional setting, or
higher-dimensional for that matter, if the basic objects are generalized.

The appearance of the Riemann zeta function  in  the effectively one-dimensional setting (cf.\ for example \cite{Sachdev}, where an integral with Jacobi theta like the one above was found as the cylinder limit of  a torus)
can be thought of as a prototype for L-functions associated with modular forms  (see e.g.\ \cite{Bump}, section 1.1). 
It is in general an interesting question whether one can recover a modular form from a given L-function (a ``converse theorem''), 
similarly to how here, the Jacobi theta function was recovered from an inverse Mellin transformation.
 
More physically, this kind of method can be used to go beyond perturbation theory in the mass parameter $\mu$. The integral
in \eqref{Mellt} has exponential suppression from the mass term, but this suppression 
is lost when we study specific terms in a series expansion in $\mu$. 

One could for example compute correlation functions in the deformed
theory in terms of massive Kronecker-Eisenstein series.  An interesting setting for this is vector-valued modular forms, which have been
used to efficiently compute physical correlators in the Ising model on the torus \cite{Cheng:2020srs}. 
Another example is that one can consider the large-mass limit,
which in \cite{Duplantier:1987sd} is interpreted as a melting process. 
Finally, it would be a natural continuation in the spirit of the paper \cite{Itzykson1989}, which inspired this work, 
  to obtain a complete expression for the effective central charge including magnetic field, to see its evolution toward the Lee-Yang edge singularity in imaginary magnetic field.

\section*{Acknowledgements}
I would like to thank Nordita for warm hospitality while this work was being performed.
I thank Michael Haack, Joseph Maciejko, Sergej Moroz and Kostya Zarembo for useful comments on the manuscript. 
I also thank Max Downing, Sameer Murthy, and G\'erard Watts for sharing their very nice work \cite{Downing:2023uuc} before publication,
and for useful comments on my manuscript. I also thank the anonymous referees for many useful comments.

\appendix

\section{The ordinary (massless) Kronecker-Eisenstein series}
\label{class}
The ordinary (massless) Kronecker-Eisenstein series is
\be \label{Em0}
E_s(z,\tau)=
\pi^{-s} \sum'_{c,d} {\tau_2^s e^{2\pi i \theta_{c,d}} \over |c\tau + d|^{2s}}
\ee
where $z = x + \tau y$ and the phase angle $\theta_{c,d} = cx - dy$ for real variables $x,y$. Explicitly, if we write real and imaginary parts
as $z=z_1+iz_2$ then $x = z_1-(\tau_1/\tau_2)z_2$ and $y =z_2/\tau_2$. The prime on the sum means to exclude the $c=d=0$ term.

\subsection{Torus translations}
Under the translation $z \rightarrow z+1$ around one of the cycles of the torus, we have $x\rightarrow \tilde{x} = x+1$ and $y\rightarrow \tilde{y}=y$ (invariant), so
$e^{2\pi i \tilde{\theta}_{a,b}} = e^{2\pi i \theta_{a,b}}$ and
\be \label{Em0trans}
E_s(z+1,\tau)=E_s(z,\tau) \; . 
\ee
Under the translation $z \rightarrow z+\tau$ around the other cycle of the torus, we have $x\rightarrow \tilde{x} = x+\tau_1-(\tau_1/\tau_2)\tau_2=x$ (invariant)
and $y \rightarrow \tilde{y}=y+1$,
so again $e^{2\pi i \tilde{\theta}_{a,b}} = e^{2\pi i \theta_{a,b}}$:
\be
E_s(z+\tau,\tau)=E_s(z,\tau) \; . 
\ee

\subsection{Modular transformations}
Under the S transformation $\tau \rightarrow -1/\tau$, we have $\tau_ 2\rightarrow \tau_2/|\tau|^2$. 
Take $z\rightarrow z/\tau$, then we have
\be \label{Em0S}
E_s(z/\tau,-1/\tau)=
\pi^{-s} \sum'_{c,d} {(\tau_2/|\tau|^2)^s e^{2\pi i \tilde{\theta}_{c,d}} \over |c(-1/\tau) + d|^{2s}} = E_s(z,\tau) \; , 
\ee
by relabelling $c\rightarrow d$, $d\rightarrow -c$ and identifying $\tilde{x} = y$, $\tilde{y}=-x$, for which
the marked point is $\tilde{z} = \tilde{x} + \tilde{\tau}
 \tilde{y}$ in the dual (reciprocal) lattice.
  
 Similarly, under  the T transformation  $\tau \rightarrow \tau+1$, we have
\be \label{Em0T}
E_s(z,\tau+1)=
\pi^{-s} \sum'_{c,d} {\tau_2^s e^{2\pi i \tilde{\theta}_{c,d}} \over |c(\tau+1) + d|^{2s}} = E_s(z,\tau) \; ,
\ee
by relabelling $d\rightarrow d -c$ and identifying $\tilde{x} = x-y$, $\tilde{y}=y$, for which $\tilde{z} = \tilde{x} + \tilde{\tau} \tilde{y}$.

Now, view $E_s(z,\tau)$ for $z=1/2$, $z=\tau/2$, $z=1/2+\tau/2$ as the three even spin structures: topological sectors of quasiperiodicities for the fermions.
(For a geometric argument why that is so, see \ref{Ejacobi}.)
Since  S acts on $z$, we recover the standard statement that the partition function for each topological sector is
 not separately modular invariant, i.e.\ invariant under both S and T. 
 But if we add the three pieces, and use torus translations
$z \rightarrow z+1$ and $z\rightarrow z+\tau$,
we see that the sum of all three sectors is modular invariant. We are then free to manipulate the pieces separately, with the understanding
that at the end we sum over topological sectors. 

\subsection{Weyl reflection: $s\rightarrow 1-s$}
\label{Weylrefl}
The Kronecker-Eisenstein series has  a functional relation 
much like the reflection formula for the Riemann zeta function. 
Under reflection $s\rightarrow 1-s$, the Kronecker-Eisenstein series transforms as
\be \label{refl}
E_{s}(z,\tau) = {\Gamma(1-s) \over \Gamma(s)} \widetilde{E}_{1-s}(z,\tau)
\ee
where in the reflected $\widetilde{E}_s(z,\tau)$, the phase in \eqref{Em0} is moved to a shift $z$ in the denominator:
\be \label{Edual}
\widetilde{E}_s(z,\tau)=
\pi^{-s} \sum_{c,d} {\tau_2^s  \over |z + c\tau + d|^{2s}} \;. 
\ee
In the notation 
of Siegel's Tata lectures \cite{siegel} (reproduced in Appendix E of \cite{Berg:2014ama}) the Kronecker-Eisenstein series is written
as a more general function $E_s(w,z,\tau)$ of three variables, whereas here one of them is always zero: 
$E_s(z,\tau)=E_s(0,z,\tau)$ and $\widetilde{E}_s(z,\tau) = E_s(z,0,\tau)$.

For $s\rightarrow 0^-$ the $1/\Gamma(s)$ factor in  \eqref{refl} has a zero, 
which tells us that only a pole at $s=0$ of $ \widetilde{E}_{1-s}(z,\tau)
$ can contribute. So $E_0$ and $E_0'$ (derivative with respect to $s$) are determined from  \eqref{refl}, as:
\be
E_0 + s E_0' + \ldots = \left(s + 2\gamma s^2 + \ldots \right) \left(-{1 \over s} + \widetilde{E}_1^{\rm reg}+\ldots\right)
\ee
where the dots mean higher order in $s$. This fixes
\be \label{E0vals}
E_0(z) \equiv -1 \; , \quad
E_0'(z) = \widetilde{E}_1^{\rm reg}(z)-2\gamma
\ee
where the regularized $ \widetilde{E}_1^{\rm reg}(z)$ is
\be \label{Elim}
 \widetilde{E}_1^{\rm reg}(z) = \lim_{s\rightarrow 1^+}  \left( \widetilde{E}_s(z) -{1 \over s-1} \right) \; . 
\ee
It is not easy to extract reliable numerical values for  $ \widetilde{E}_1^{\rm reg}(z)$ from a limiting procedure like \eqref{Elim}, since the double sums converge slower
and slower for $s\rightarrow 1^+$. This is not a problem for correlation functions, where also the third argument is nonzero, in which case there is no pole as $s\rightarrow 1$. 

To put this in context, the reflection $s \rightarrow 1-s$ is a special case of a  Weyl reflection on a Lie algebra root lattice, see e.g.\ \cite{Fleig:2015vky}, section 10.3. 

\subsection{Kronecker-Eisenstein series as Jacobi theta functions}
\label{Ejacobi}
\begin{figure}[h]
\begin{center}
\includegraphics[width=0.5\textwidth]{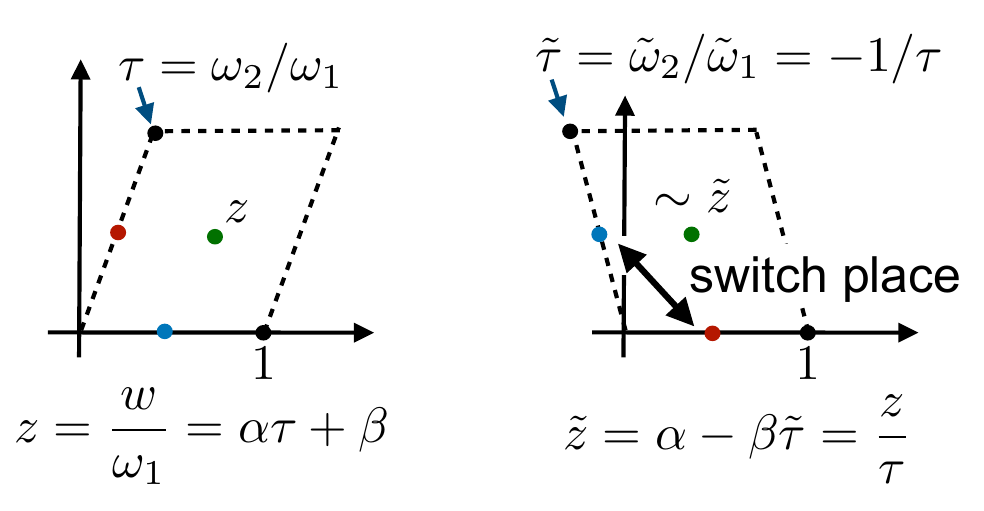}
\caption{The modular transformation that switches the lattice basis vectors (cf.\  fig.\ \ref{fig:ztr}) switches the half-periods $z=1/2$ and $z=\tau/2$ (where $\vartheta_2$, $\vartheta_4$ have zeros) but
keeps $z=0$ and $z=1/2+\tau/2$  (where $\vartheta_1$, $\vartheta_3$ have zeros) at the corresponding place in the (dual) torus lattice labelled by $\tilde{\tau}$ .}
\label{fig:Str}
\end{center}
\end{figure}

Kronecker's second limit formula from 1863 \cite{Kronecker} states that for $z\neq 0$,
\be \label{E1theta}
E_1(z) = -\ln\left| {\vartheta_1(z,\tau) \over \eta(\tau)}\right|^2 + {2\pi z_2^2 \over \tau_2}
\ee
where $z_2=$ Im $z$ and $\vartheta_1$ is a Jacobi theta function. Exponentiating \eqref{E1theta}, evidently
\be
e^{-E_1(z)} = e^{-2\pi z_2^2/\tau_2} \left| {\vartheta_1(z,\tau) \over \eta(\tau)}\right|^2 \; ,
\ee
where the factor in absolute value can be recognized as a partition function.
The (${\mathbb Z}_2$) Jacobi theta functions $\vartheta_1$, $\vartheta_2$, $\vartheta_3$, $\vartheta_4$ can be distinguished by where they have a zero. It then follows from 
the geometry in fig.\ \ref{fig:Str} that $\vartheta_2$ and  $\vartheta_4$ are switched by the S modular transformation.

Now consider the four ${\mathbb Z}_2$ fermion quasiperiodicities  around the two cycles of the torus: even or odd.
Switching the axes, (odd, even) and (even, odd) are exchanged. 
To represent the partition function with the complex number $z=\alpha \tau + \beta$ for $\alpha,\beta=0,1/2$,
i.e.\ one of the four marked points in fig.\ \ref{fig:Str}, means to relate the zeros of Jacobi theta functions to fermion quasiperiodicities $e^{2\pi i \alpha}$, $e^{2\pi i \beta}=\pm 1 $ (even/odd).  

This simple argument is supposed to illustrate that here, the marked point and the quasiperiodicities represent the same thing, but in general, it is not obvious a priori:  theta functions in the partition function arise by solving a local differential equation to compute a functional determinant, whereas global quasiperiodicity is a topological property. An index theorem
relates these local and global properties. For more on this, see e.g.\ \cite{Fleig:2015vky} section 13.4. 

\subsection{Inversion in $\mu$ and reflection in $s$}
\label{invmu}
In \cite{Cardy:2022mhn} it was pointed out that a mass deformation modifies the infrared spectrum (low $n$ in $q^n$).
Although this is true, there is also a certain duality relation $\mu \leftrightarrow 1/\mu$. 
For $w\neq 0$, $z\neq 0$ (which is not the main case we consider here, but for the purposes in this section,
it is convenient to have fewer poles),
we have the reflection 
\be \label{Erefl}
E_s(w,z,\tau) = e^{{2\pi i \over \tau_2} {\rm Im}(w\overline{z})} \widetilde{E}_{1-s}(z,w,\tau)
\ee
(actually here the tilde is not needed, but it will be useful below)
and the massive Kronecker-Eisenstein series is given by an inverse Mellin transformation similarly to the $w=0$ case
that is the main focus in this paper:
\be \label{moregeneralE}
\mathcal E_{1,\mu}(w,z,\tau) := \int_{c-i\infty}^{c+i\infty} (\pi \mu)^{-s} \Gamma(s)E_{s+1}(w,z,\tau) ds \; . 
\ee
For brevity, in the following equation  $z$ and $w$ are suppressed, but they are still nonzero to avoid additional poles.
The massive Kronecker-Eisenstein series for the inverse mass $\mu^{-1}$ is:
\bea \label{mudual}
 \pi^{2s}\mathcal E_{1,\mu^{-1}}(\tau) &=&\int_{c-i\infty}^{c+i\infty} (\pi \mu)^{s} \Gamma(s)E_{s+1}(\tau) ds
=\xi  \int_{c-i\infty}^{c+i\infty} (\pi \mu)^{s} {\Gamma(s)\Gamma(-s)\over \Gamma(s+1)} \widetilde{E}_{-s}(\tau) ds \nonumber \\
&=&
\xi  (\pi \mu)^{-1} \int_{-1-c-i\infty}^{-1-c+i\infty} (\pi \mu)^{-\tilde{s}} {\tilde{s}\over  \tilde{s}+1}\Gamma(\tilde{s}) E_{\tilde{s}+1}(\tau) d\tilde{s}
\eea
where $\tilde{s} = -1-s$ and $\xi = e^{{2\pi i \over \tau_2} {\rm Im}(w\overline{z})}$. 
%All poles of this expression except $\tilde{s}=-1$ now occur in the positive  half-plane Re $\tilde{s}>0$,
%where the double sum representation of $E_{\tilde{s}+1}$ converges, and at $\tilde{s}=-1$  the residue
%is $E_{0}\equiv 0$, if $w\neq 0$. 
Since we already had  $\mathcal E_{1,\mu}(\tau)$ in a form like this  in \eqref{moregeneralE}, 
the rewriting \eqref{mudual} gives a duality relation $\mathcal E_{1,\mu^{-1}}(\tau)\leftrightarrow \mathcal E_{1,\mu}(\tau)$. (Note ``relation'': it
is in general not an identity, just a mapping.) We will not explore this further here.

In \cite{Kostov:2022pvy}, the high- and low-temperature expansions were compared, and it was argued that the sign of $D_{0,0}$ in \eqref{Ising1}
should be negative for $T>T_{\rm c}$. The $D_{0,0}$ piece does not affect most of the discussion in this paper, but the methods
of \cite{Kostov:2022pvy}
may be useful for developing arguments as in this section for  high- vs.\  low-temperature expansions, viewed
as maps of $\mu$ similar to $\mu\rightarrow \mu^{-1}$ considered here.\footnote{I thank G\'erard Watts for alerting me to this.}

\section{Basics of Bessel sums}
\label{besselapp}
It is somewhat instructive to compute  Bessel sum like that in \eqref{cexpl} by elementary means.
The Bessel-$K_0$ cosine series can be computed using the integral representation \href{https://dlmf.nist.gov/10.32#E6}{DLMF 10.32.6} \cite{DLMF}
\be \label{DLMF1}
K_0(z) = \int_0^{\infty} \!dy {\cos (zy) \over \sqrt{y^2+1}}  = \int_0^{\infty} \!dy {y\sin (zy) \over z(y^2+1)^{3/2}} \; . 
\ee
where  for later purposes, the second version increased the power in the denominator by integration by parts.

\begin{figure}[h]
\begin{center}
\includegraphics[width=0.5\textwidth]{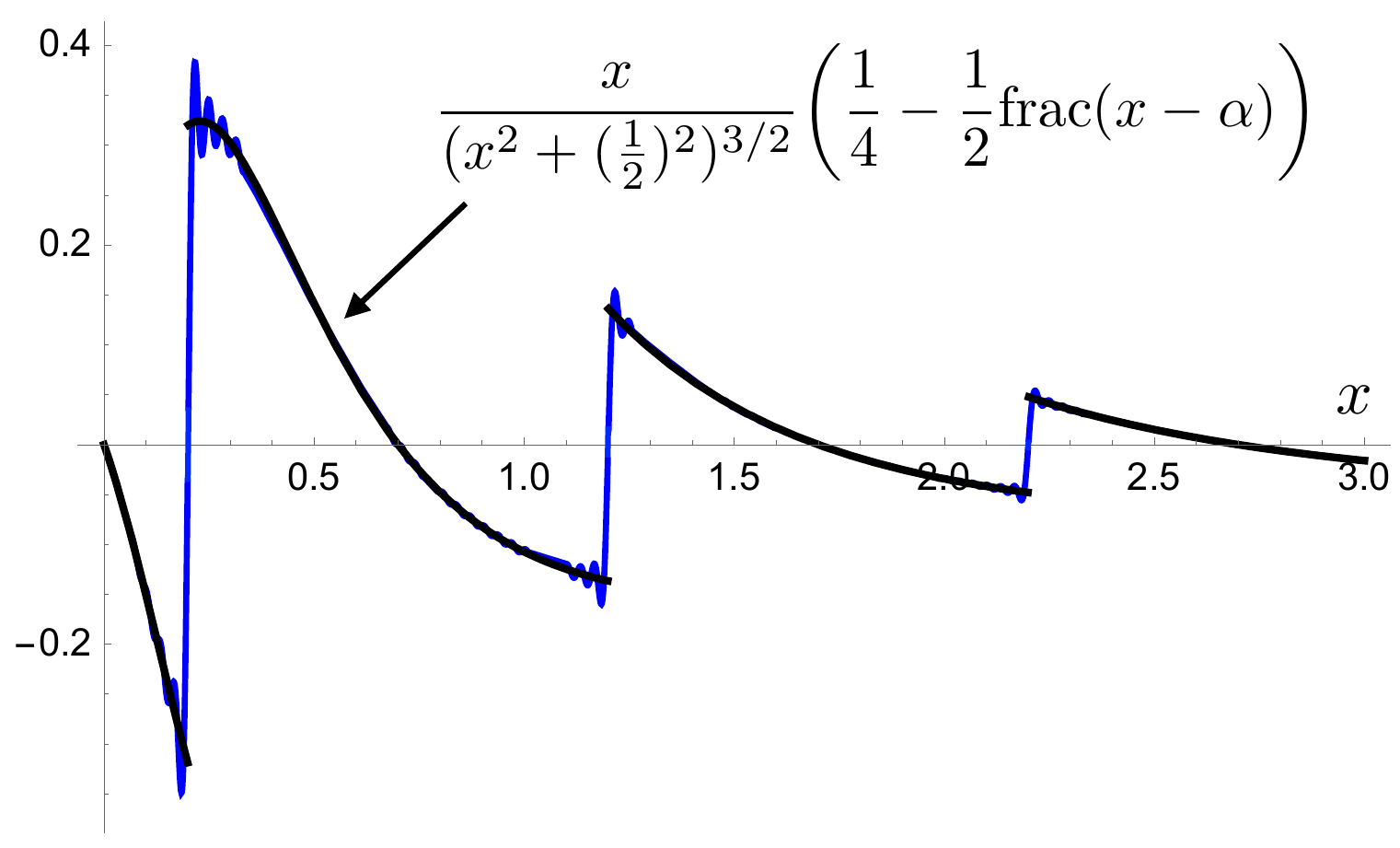}
\caption{The integrand of \eqref{plotsaw}, times 2, for $m=1/2$, $\alpha=0.2$ (black),
and the corresponding Fourier representation \eqref{Four} truncated to 30 terms (blue). Note that
since there is suppression in $x$, the sawtooths do not integrate to zero.}
\label{fig:saw}
\end{center}
\end{figure}

Setting $z=2\pi n m$,  we can change variables $my = x$:
\be
K_0(2\pi n m) = \int_0^{\infty} \!  dy {y\sin (2\pi n m y) \over 2\pi n m(y^2+1)^{3/2}} = 
 \int_0^{\infty} \!{x\sin (2\pi n x) \over 2\pi n (x^2+m^2)^{3/2}}\,  dx   \; . 
\ee
Multiplying by $\cos(2\pi n \alpha)$ and summing over $n$ we have
\be
\sum_{n=-\infty}^{\infty}K_0(2\pi n m) \cos(2\pi n \alpha)= \sum_{n=-\infty}^{\infty}\int_0^{\infty} \! dx {x\sin (2\pi n x) \over 2\pi n (x^2+m^2)^{3/2}}
\cos(2\pi n \alpha) 
\ee
Now combine $\sin(2\pi n x)\cos(2\pi n \alpha) = {1 \over 2}(\sin(2\pi n(x-\alpha))+\sin(2\pi n(x+\alpha))$:
\be \label{sumsaw}
\sum_{n=-\infty}^{\infty}\int_0^{\infty} \! dx\, {x \over (x^2+m^2)^{3/2}}
{1 \over 2}\left({\sin(2\pi n(x-\alpha))\over 2\pi n}+{\sin(2\pi n(x+\alpha)\over  2\pi n} \right) \; . 
\ee
Each term is no longer a Fourier cosine series,
but a sawtooth wave:
\be \label{Four}
\sum_{n=1}^{\infty} {\sin (2\pi n y) \over 2\pi n} = {1 \over 4}-{1 \over 2}{\rm frac}(y) \; ,
\ee
so \eqref{sumsaw} can be written with the sum already performed:
\be \label{plotsaw}
\int_0^{\infty} \!dx \, {x \over 2(x^2+m^2)^{3/2}}\left({1 \over 4}-{1 \over 2}{\rm frac}(x-\alpha)+
{1 \over 4}-{1 \over 2}{\rm frac}(x+\alpha)\right)
\ee
where the integrand is plotted in fig.\ \ref{fig:saw}. 
If we split the integral into each sawtooth in fig.\ \ref{fig:saw}, each piece is elementary to perform.
For example, for $\alpha=0$, we find for the interval $0<x<1$ that
\be
\int_0^{1}{x \over 2(x^2+m^2)^{3/2}}\left({1 \over 2}-x\right)dx = {1 \over 4\sqrt{1+m^2}}
+{1 \over 4m} -{1 \over 2}{\rm arccoth} \sqrt{1+m^2} \; ,
\ee
so adding all the sawtooths can be viewed as the origin of the  square root sums like in \eqref{cexpl} in the main text, recalling that the standard integral representation  in eq.\ \eqref{DLMF1} was for the $K_0$ Bessel function.

 For general $K_{\nu}$, also with some power of $m$ in front, we replace the standard integral representation \eqref{DLMF1} with for example
\be
{m^{\nu} \over 2\pi n} K_{\nu}(2\pi n m ) = {\pi^{-\nu-5/2}
\Gamma(\nu+{\scriptstyle{3 \over 2}})\over n^{\nu+2}} \int_0^{\infty}\!
dx \,  {x \sin (2\pi m n x) \over (x^2+1)^{\nu+3/2}} \; . 
\ee

 Leaving integral representations, a  more direct way to obtain the square root sums in eq.\ \eqref{cexpl}  is to use Poisson summation,
as by Watson in \cite{Watson}, and reproduced and generalized in \cite{Berndt2}:
\be
{\scriptstyle {1 \over 2}}\Gamma(\nu)
+2\sum_{n=1}^{\infty}({\scriptstyle {1 \over 2}}nz)^{\nu}K_{\nu}(z)
=\Gamma({\scriptstyle {1 \over 2}})\Gamma(\nu+{\scriptstyle {1 \over 2}}) z^{2\nu}
\left({1 \over z^{2\nu+1}}+2\sum_{n=1}^{\infty}
{1 \over (z^2+4n^2\pi^2)^{\nu+{\scriptstyle {1 \over 2}}}}\right) \; . 
\ee
This formula is only valid for Re $\nu>0$, but Watson explains
how to go to $\nu\rightarrow -1$ by analytic continuation. We see that for $\nu\rightarrow -1$, indeed the corresponding
sum for the $K_1$ Bessel function will involve
a sum over terms like  $\sqrt{z^2+4n^2\pi^2}$, as in eq.\ \eqref{cexpl} from which we began.

%Expanding the Bessel function for general $\nu$ we get two terms corresponding to $I_{\nu}$ and $I_{-\nu}$:
%\begin{eqnarray} \label{Li}
%\sum_n m K(2\pi n m) {\cos(2\pi n \alpha) \over 2\pi n}&=&\sum_{n,p} \left({(2\pi n m)^p\over \Gamma(p)} -{(2\pi n m)^p\over \Gamma(p)}\right) {\cos(2\pi n \alpha) \over 2\pi n}\\
%&=& \sum_{p}\left({\rm Li}_{-2p} +{\rm Li}_{-2p} \right)m^p
%\end{eqnarray}
%where the polylogarithms have their standard analytic continuation to negative order.
%From this point of view, the four terms in the integral representation arise as two terms from $\cos x = (e^{ix}+e^{-ix})/2$, times two terms
%from the $I_{\nu}$ and $I_{-\nu}$ series.

The elementary manipulations above  fit into a greater context as follows.
If we know how an automorphic form transforms,
this produces a summation identity for the Fourier sum. 
For example, if we Fourier expand the identity ${E}_s(\tau_2)=E_s(1/\tau_2)$
we find an identity that according to \cite{Cohen} was given by Ramanujan:
\begin{eqnarray} \label{raman}
&& \hspace{-2cm} 4\sqrt{x}\sum_{n=1}^{\infty}{\sigma_s(n) \over n^{s/2}}K_{s/2}(2\pi n x)=  
{4 \over \sqrt{x}}\sum_{n=1}^{\infty}{\sigma_s(n) \over n^{s/2}}K_{s/2}\left({2\pi n \over x}\right)\\[0mm]
 &&+\xi(-s)(x^{-(1+s)/2}-x^{(1+s)/2})-  \xi(s)(x^{(1-s)/2}-x^{(s-1)/2}) \nonumber
\end{eqnarray}
where $\sigma$ is the divisor function and $\xi(s)=\pi^{-s/2}\Gamma(s/2)\zeta(s)$. 
For $s=1$, when half-integer Bessel functions reduce
to exponential functions, implies the modular transformation of the Dedekind function $|\eta(\tau)|$,
that occurs in the partition function in \ref{Ejacobi}.

\section{Conformal perturbation theory}
\label{pert}

The strategy in conformal perturbation theory is to integrate over correlation functions. Here I exhibit the two-point case,
but it is important to pick a strategy that can be pursued to higher order.
We need to integrate:
\be
{1 \over |x_2-x_1|^{4h}}
\ee
over disks with $|x_i|<\rho$.  Following \cite{Saleur1987}, 
we can write $x_1=r_1 e^{i\theta_1}$, $x_2=r_2 e^{i\theta_2}$, and $\theta=\theta_1-\theta_2$:
\be
{1 \over |r_2 e^{i\theta_2}-r_1 e^{i\theta_1}|^{4h}} = 
{1 \over r_2^{2h}|1-(r_1/r_2) e^{i\theta}|^{4h}} = {1 \over  r_2^{2h} (1-2(r_1/r_2)\cos\theta+(r_1/r_2)^2)^{2h}}
\ee
An elementary  way to perform the angle integral is to expand in Gegenbauer (ultraspherical) polynomials $C_n^{(\lambda)}(x)$, for example 
DLMF  \cite{DLMF} (18.12.4) or  (8.930) in  \cite{GR} gives the generating function:
\be
{1 \over (1-2xz +z^2)^{\lambda}}  = \sum_{n=0}^{\infty} C_n^{(\lambda)}(x) z^n
\ee
where we have $x=\cos \theta$ and $\lambda=2h$. 
An explicit representation as a finite sum is for example DLMF \cite{DLMF} (18.5.11):
\be \label{Geg}
C_n^{(\lambda)}(\cos\theta) =\sum_{\ell=0}^n {(\lambda)_{\ell}(\lambda)_{n-\ell} \over \ell! (n-\ell)!}\cos((n-2\ell)\theta)
\ee
where the Pochhammer symbol is
\be
(\lambda)_n = {\Gamma(\lambda + n) \over \Gamma(\lambda)} \;. 
\ee
Integrating eq.\ \eqref{Geg} over $\theta$  gives zero for odd $n$, but for even $n$, it gives nonzero for the ``middle'' term $\ell=n/2$,
which is in fact constant (i.e.\ independent of $\theta$). Setting $n=2m$, we have
\be
\int_0^{2\pi}{d\theta \over 2\pi}\sum_{\ell=0}^n {(\lambda)_{\ell}(\lambda)_{n-\ell} \over \ell! (n-\ell)!}\cos((n-2\ell)\theta)
=\left({\Gamma(m+\lambda) \over m! \, \Gamma(\lambda) }\right)^2 
\ee
with $\lambda=2h$. This is what we need to show eq.\ \eqref{C2general}.

\end{document}